\documentclass[a4paper,fleqn]{cas-sc}

\usepackage[authoryear]{natbib}
\usepackage{bm}
\usepackage{float}
\usepackage{wrapfig}
\floatstyle{plaintop}
\restylefloat{table}
\usepackage[section]{placeins}
\usepackage[subrefformat=parens,labelformat=parens]{subfig}
\usepackage{amsmath}
\usepackage{amsthm}
\usepackage{subfig}
\usepackage{graphicx}
\usepackage{setspace}
\usepackage{multicol}
\usepackage[ruled,vlined]{algorithm2e}
\newtheorem{defi}{Definition}
\newtheorem{prop}{Proposition}
\newcolumntype{O}{>{\centering\arraybackslash}m{1.6cm}}
\newcolumntype{M}{>{\centering\arraybackslash}m{1.8cm}}
\newcolumntype{N}{>{\centering\arraybackslash}m{0.6cm}}
\usepackage{xcolor}
\usepackage{amssymb}
\usepackage{comment}

\def\tsc#1{\csdef{#1}{\textsc{\lowercase{#1}}\xspace}}
\tsc{WGM}
\tsc{QE}
\tsc{EP}
\tsc{PMS}
\tsc{BEC}
\tsc{DE}

\begin{document}
\def\floatpagepagefraction{1}
\def\textpagefraction{.001}
\shorttitle{T. Lilasathapornkit et~al.}
\shortauthors{T. Lilasathapornkit et~al.}

\title [mode = title]{Traffic Assignment Problem for Footpath Networks with Bidirectional Links}

\author[1]{Tanapon Lilasathapornkit}


\address[1]{Research Centre for Integrated Transport Innovation (rCITI), School of Civil and Environmental Engineering, University of New South Wales, Sydney, NSW, Australia}

\author[1,2]{David Rey}

\address[2]{SKEMA Business School, Sophia Antipolis, France}

\author[1,3]{Wei Liu}

\address[3]{School of Computer Science and Engineering, University of New South Wales, Sydney, NSW, Australia}

\author[1]{Meead Saberi}
\cormark[1]

\cortext[cor1]{Corresponding author: meead.saberi@unsw.edu.au}

\begin{abstract}
The estimation of pedestrian traffic in urban areas is often performed with computationally intensive microscopic models that usually suffer from scalability issues in large-scale footpath networks. In this study, we present a new macroscopic user equilibrium traffic assignment problem (UE-pTAP) framework for pedestrian networks while taking into account fundamental microscopic properties such as self-organization in bidirectional streams and stochastic walking travel times. We propose four different types of pedestrian volume-delay functions (pVDFs), calibrate them with empirical data, and discuss their implications on the existence and uniqueness of the traffic assignment solution. We demonstrate the applicability of the developed UE-pTAP framework in a small network as well as a large scale network of Sydney footpaths.

\end{abstract}


\maketitle

\section{Introduction}

Rapid urban population growth in the past few decades has made understanding and predicting pedestrian traffic an increasing challenge in major city centers across the world with overcrowded footpaths. The growing interest in improving walkability and the need for more comprehensive appraisal of walking infrastructure in cities require reasonably accurate estimates of pedestrian traffic volumes. However, the modeling methodologies and tools to estimate foot traffic have been an overlooked area of research in the literature. {\color{black}To the best of our knowledge, no study has explicitly extended the traffic assignment problem to pedestrian networks in the urban context taking into account the walking route choice and microscopic behaviour of pedestrian crowds such as self-organization and formation of lanes.}

The traffic assignment problem (TAP) has been subject to intense research since the 1950's building upon seminal studies of \citet{wardrop1952road}, \citet{beckmann1956studies}, and \citet{smith1984stability}. However, almost all studies on TAP have focused on car traffic in which vehicle flow on a link is associated with travel time on the same link known as the link performance function or volume-delay function (VDF) \citep{bureau1964traffic}. A common assumption in the TAP is that the link performance functions are independent of each other, making the travel time on a given link depend only on the flow through that link and not on the flow through any other link in the network. Link flows interaction was first introduced in the TAP for heavy traffic on two-way streets, signalized intersections, merging sections, left-turning movements in unsignalized intersections, and multi-modal traffic using a pairwise symmetric link interaction function \citep{Sheffi1985, abdulaal1979methods, dafermos1971extended, chevallier2007macroscopic}. \citet{beckmann1956studies} showed that the TAP has a unique User Equilibrium (UE) solution if the link travel time function is strictly monotonic with respect to its link traffic flow. For the UE solution to remain unique when link flow interactions exist, the flow in the main direction must affect travel time more than the flow in the counter direction \citep{Sheffi1985}.

{\color{black}Unlike car traffic, pedestrian traffic does not often follow pre-specified lanes. Links in a pedestrian network (e.g. footpaths) accommodate bidirectional flow \citep{tordeux2018mesoscopic}.} Therefore, pedestrian travel time in one direction depends on the flow of the same direction as well as the flow of the counter direction. When the flow in the counter direction is high, pedestrians in the main direction have less room to freely walk at their desired speed regardless of the flow in the main direction. Therefore, the effective capacity and average speed of the main direction reduces as the flow of the counter direction increases \citep{lam2003generalised,kretz2006experimental,Guo2012a, feliciani2018universal,taherifar2019macroscopic}. Research on pedestrian dynamics has grown in several directions in the past decade including a deeper exploration of pedestrian fundamental diagram \citep{Floetteroed2015,nikolic2016probabilistic,hoogendoorn2018macroscopic}, microscopic modeling \citep{huang2017behavior,tao2017cellular,shahhoseini2018pedestrian}, mesoscopic modeling \citep{tordeux2018mesoscopic}, and macroscopic simulation \citep{hanseler2017dynamic,taherifar2019macroscopic,aghamohammadi2018dynamic,moustaid2021macroscopic}. Several studies in the literature have studied bidirectional pedestrian dynamics using the fundamental relationship between flow and density \citep{Seyfried2005,zhang2012ordering,hanseler2014macroscopic,cao2017fundamental,hanseler2017dynamic}. Despite its significance and practical relevance, very little effort has been put into understanding the network-wide impact of pedestrians in the urban context for planning applications. This knowledge gap is the main objective and motivation of this research. To the best of authors' knowledge, bidirectional link performance function in the context of pedestrian TAP has not been studied before.

In this paper, we propose a TAP framework for pedestrian networks, termed pedestrian TAP (pTAP). We present new link performance function forms specifically calibrated for bidirectional pedestrian streams, representing two-way footpaths in urban centers, termed pedestrian VDF (pVDF). We propose a UE-pTAP formulation with a stochastic symmetric pVDF under the assumption that travel times of the two opposing directions are equal on the same bidirectional stream. We extend the formulation to UE-pTAP with deterministic and stochastic asymmetric pVDF assuming that travel times of the two opposing directions may no longer be equal. The deterministic pVDF captures the impact of bidirectional flow while the stochastic pVDF accounts for dynamic and unstable nature of the self-organization phenomena resulting in high variability of travel time for any given flow. Formation of self-organized lanes in overcrowded environments often improves the speed of pedestrians \citep{Saberi2014} and the capacity of the walkway \citep{feliciani2018universal}. Pedestrian dynamics can be  greatly affected by formation of self-organized lanes \citep{hoogendoorn2015continuum, haghani2018crowd} and the capacity of the walking corridor \citep{shahhoseini2018pedestrian}. In this study, we use empirical data from \citep{zhang2012ordering, zhange2012phd,holl2018methoden} to calibrate the proposed pVDFs. We apply the proposed UE-pTAP framework on a small toy network and the Sydney Central Business District (CBD) footpath network.

The remainder of the paper is organized as follows. A review of the literature is provided in Section \ref{sec:lit}. In section \ref{sec:method}, we present the methodological contributions including the walking network representation, deterministic and stochastic  bidirectional pVDFs, and the proposed UE-pTAP formulations. Section \ref{sec:calibration} presents the calibration results of the proposed pVDFs against controlled experimental data. Section \ref{sec:case} presents the numerical results. Section \ref{sec:conclusion} summarizes the contributions and suggests future research directions.

\section{Literature Review} \label{sec:lit}
\subsection{Traffic assignment problem for vehicular traffic}
Traffic assignment's primary objective is to provide link and network level performance measures such as link volumes and total system travel time \citep{ortuzar2011modelling}. TAP is widely used as a part of major transportation infrastructure projects appraisal. Traffic assignment is also used to estimate turning movements for intersection design, traffic pollutant emission, and traffic noise \citep{aumond2018probabilistic, amorim2013pedestrian}. The main inputs to the model are the structure of the network, travel demand, and the link cost functions. To facilitate readability of the paper, a nomenclature is provided in Appendix~\ref{sec:appendix_notation}. Formally, we denote $G=(N,A)$ the directed graph consisting of nodes $N$ and links $A$. Let $q_{rs}$ be the travel demand from origin node $r \in N$ to destination node $s \in N$. We denote $W$ the set of origin-destination (OD) pairs in the network, i.e. $W = \{(r,s) : r \in N, s \in N, q_{rs} > 0\}$ and denote $\Pi_{rs}$ the set of paths in $G$ connecting OD pair $(r,s) \in W$. Let $\left[\delta^{rs}_{a,k}\right]$ be the link-path incidence matrix defined as
\begin{equation} \color{black}
\delta^{rs}_{a,k} = \begin{cases}1 & \textrm{if link $a$ is on path $k$ of OD $(r,s)$}\\ 0&\textrm{otherwise} \end{cases} \qquad \forall a \in A, \forall (r,s) \in W, \forall k \in \Pi_{rs}. 
\label{eq:incident}
\end{equation}

Let $\mathbf{t}$ be the vector of link travel times, $\mathbf{x}$ be the vector of link flows, $\mathbf{x}^{*}$ be the optimized vector of link flow, $x_a$ be the link flow on link $a \in A$, $t_{a}(x_a)$ be the travel time on link $a \in A$ as a function of the link flow $x_{a}$, and $f^{rs}_{k}$ be  flow on path $k$ connecting OD pair $(r,s) \in W$. \citet{wardrop1952road}'s first principle states that at user equilibrium (UE), no traveler has any incentive to unilaterally change its route. {\color{black}The well-known Beckmann formulation represents the TAP under UE conditions \citep{beckmann1956studies} resulting from minimizing a nonlinear convex function under the condition that the Jacobian of the link travel times has to be symmetric \citep{nagurney1993network}. To circumvent this limitation, the Variational Inequality (VI) formulation is alternatively used that can handle an asymmetric Jacobian of link travel times \citep{smith1979existence, dafermos1980traffic, bertsekas1982projection, lawphongpanich1984simplical,bliemer2001analytical,han2004solving}.} Let $\Omega$ defined as the set of all feasible link flows \textbf{x} 

\begin{equation}
\Omega = \Big \{ \mathbf{x} \geq 0 : x_{a} = \sum_{r \in N} \sum_{s \in N} \sum_{k \in \Pi_{rs}} \delta^{rs}_{a,k} f^{rs}_{k}, \forall a \in A, \sum_{k \in \Pi_{rs}} f^{rs}_{k} = q_{rs}, \forall (r,s) \in W, f^{rs}_{k} \geq 0, \forall (r,s) \in W, \forall k \in \Pi_{rs} \Big \}
\label{eq:wardrop1st_constraint}
\end{equation}

The VI formulation seeks to find a set of link flows $\mathbf{x}^{*}$ as expressed in Equation~\eqref{eq:VI} \citep{mannering2013principles,ortuzar2011modelling,tfl2010traffic,roads2013traffic,hoogendoorn2018macroscopic}

\begin{equation}
\hspace{120pt} \textbf{t} ^{\mathsmaller T} (\textbf{x}^{*})(\textbf{x} - \textbf{x}^{*}) \geq 0 \hspace{30pt} \forall \textbf{x} \in \Omega \label{eq:VI} 
\end{equation}

In this paper, we focus on Wardrop's first principle since it represents the general traffic condition that all travelers seek to serve their best interests. UE models may simplify reality considerably, but their efficiency, tractability, and stability make them very useful in practice.

{\color{black}The UE model is formulated mathematically to minimize a nonlinear convex function \citep{beckmann1956studies} subject to a set of linear constraints as shown in Equation \ref{eq:wardrop1st_constraint}.} Existence and uniqueness of the UE solution are important properties of the TAP. For the existence property, Weierstrass' theorem ensures that a continuous function is able to attain its minimum on a bounded solution which is a UE solution \citep{PatrikssonMichael1994Ttap}. Similarly the equivalence between the UE condition and its first-order optimal conditions ensures that the UE solution is satisfied at any local minimum \citep{Sheffi1985}. For the uniqueness property, the equilibrium travel times and link flows are from a unique global optimal solution if the objective function is strictly convex \citep{PatrikssonMichael1994Ttap, bazaraa2013nonlinear}. The objective function is strictly convex if the link travel time function is {\color{black}strictly} monotonic.

Non-monotonic link cost functions have previously been proposed and used in the literature to solve variants of the traffic assignment problem. A few examples include the use of non-monotonic emission functions \citep{benedek1998equitable,sugawara2002much,patil2016emission,yin2006internalizing,chen2012managing,tidswell2021minimising}, physical queues in dynamic traffic assignment (DTA) with non-monotonic and non-differentiable cost functions \citep{szeto2005properties}, and multiclass traffic assignment \citep{zhang2010solving}. These studies collectively provide a comprehensive discussion on the solution non-uniqueness issue emerging from using non-monotonic cost functions in the traffic assignment problem. Without a unique solution, minor changes in demand could cause the algorithm to have entirely different solutions. So comparisons between different scenarios could be unreliable and misleading. A few studies in the literature address this problem by using an algorithm to push the solution toward equilibria where the cost function is locally monotonic \citep{wynter1996solving} or introducing a weaker monotonic assumption \citep{marcotte2004new}. 

In this paper, we propose both monotonic and non-monotonic pedestrian VDFs (pVDFs) and we discuss their implications and properties when used in the traffic assignment problem. While the proposed monotonic pVDF provides neat mathematical properties for the traffic assignment solution, empirical data suggest that it may not correctly capture the pedestrian traffic dynamics. On the other hand, the proposed non-monotonic pVDF captures the bidirectional dynamics quite reasonably as supported by empirical data. {\color{black}However, it comes with its expected limitations that non-monotonic pVDF may not guarantee the solution existence and uniqueness.} Classical algorithms to determine converged equilibrium condition, also known as link-based algorithms, include  Frank-Wolfe's algorithm  \citep{frank1956algorithm,LeBlanc1975,leblanc1985improved} and the Method of Successive Average (MSA) \citep{powell1982convergence}. In this study we apply the classical MSA algorithm to solve the UE pTAP.

Link-based algorithms are known to converge very slowly. Path-based algorithms aim to improve the convergence rate  \citep{dafermos1971extended,jayakrishnan1994faster,florian2009new}. However, path-based algorithms still have drawbacks such as high computational requirements and non-unique path flows. More recently, Bush-based algorithms including Origin-Based Algorithm (OBA) \citep{bar2002origin}, Algorithm B \citep{dial2006path}, Local User Cost Equilibrium (LUCE) \citep{gentile2014local}, and Traffic Assignment by Paired Alternative Segments (TAPAS) \citep{bar2010traffic} have been introduced to further improve the convergence rate and address the uniqueness problem. {\color{black}Another type of stochastic traffic assignment model using a recursive logit model has also been widely adopted recently in vehicular networks (Bell, 1995; Fosgerau, 2013; van Oijen, 2020; Zimmermann, 2020).} In this paper, we build the pTAP upon the traditional UE formulation for TAP specifically for pedestrian networks taking into account the effect of self-organization and stochastic travel times of pedestrians.

\subsection{Link performance function}
The link performance function, also known as VDF, quantifies the travel time as a function of traffic volume on each link. Volume-to-capacity ratio reflects the traffic condition on a link while link capacity is usually expressed as the maximum number of users per unit time or as a set of maximum capacities for different levels of service and link types. The U.S. Bureau of Public Roads (BPR) \citep{bureau1964traffic}, is the most commonly used VDF in the TAP literature:

\begin{equation} \label{eq:BPRfunc}
\hspace{150pt} t\left(x\right) = \tau \left( 1+\alpha \left(\frac{x}{c}\right)^{\beta} \right), 
\end{equation}

where $\tau$ denotes free flow travel time, $c$ denotes capacity, and $x$ denotes traffic flow.

The BPR function as shown in Equation \eqref{eq:BPRfunc} has a very simple form, but also has some drawbacks. {\color{black}TAP formulations based on the BPR function may suffer from slow convergence rate due to the overestimation of travel time on congested links. It also underestimates costs of links that have low volumte-to-capacity ratio and does not provide a guaranteed unique solution if $\beta$ is large \citep{spiess1990conical}. A conical function instead is known to guarantee the uniqueness of the link volumes and does not overestimate travel time of highly congested links compared to the classical BPR function by ensuring a strictly increasing function \citep{spiess1990conical}. The VDF is often used in a deterministic form \citep{bureau1964traffic, dot1985traffic, manual2010transportation}. However, link capacities can be treated as stochastic variables following a distribution based on traffic flow patterns \citep{brilon2005reliability}. A VDF with stochastic capacity captures the probability of traffic breakdown on freeway links \citep{neuhold2014volume}. The UE assumes that all travelers have perfect information of the entire network and travelers are able to precisely predict travel times. In reality, travelers may not homogeneously react to the same traffic condition, so different traits of travelers can be categorized into risk-averse, risk-prone, and risk-neutral \citep{yin2001assessing}.} A stochastic VDF could, therefore, reflect different traveler attitudes toward risk with additional penalty upon late arrival at the end-point of a trip \citep{watling2006user}. 

Most of the proposed VDFs in the literature of the TAP only focus on a single link independently. \citep{Mueller2015} is among the very few existing studies that takes into account a type of interaction, specifically the interaction between two types of vehicles, trucks and cars. Unlike car traffic, in bidirectional pedestrian streams, travel time of the reference direction is not only affected by the flow of the reference direction, but it is also influenced by the flow from the counter direction \citep{dafermos1971extended}. Previous studies have shown that traffic flow relationships calibrated for unidirectional flows may not be appropriate to apply to bidirectional pedestrian streams \citep{taherifar2019macroscopic}. Therefore, it is essential to estimate a specific pVDF function for bidirectional pedestrian streams that more realistically mimics the performance of the foot traffic on urban footpaths.

\subsection{Bidirectional pedestrian stream}
Pedestrian motion, unlike car traffic, is not confined to dedicated lanes \citep{tordeux2018mesoscopic}. Pedestrians in bidirectional streams often intertwine with the counter direction \citep{Guo2012a} and thus, dynamically form self-organized lanes \citep{saberi2015} whereas typical vehicular traffic streams are separated by direction with no or little bidirectional interaction. Self-organized lanes can have significant impact on walking speed \citep{Saberi2014}. A recent study in the literature \citep{feliciani2018universal} demonstrated the impact of self-organized lanes on the capacity of the bidirectional streams and concluded that the flow ratio can significantly influence the capacity of a pedestrian stream. They proposed a U-shaped capacity function when self-organization does not occur and a W-shaped capacity function when dynamic lanes form \citep{feliciani2018universal}. The flow ratio represents the proportion of flow on the reference direction of interest over the total flow from both directions. A study by \citep{Wong2010} compared the asymmetric effect of the major and minor flows on speed and effective capacity when pedestrian streams walk towards each other at different angles. The study concluded that pedestrian streams walking 180 degree toward each other shows the most significant bidirectional impact. When the flow ratio of the major stream is high (e.g. above 0.8), it is expected that the major stream behaves similarly to a unidirectional stream  \citep{Lam2002}. Therefore, it is the minor stream that is significantly affected from the bidirectional effects \citep{lam2003generalised}. This occurs because pedestrians in the minor stream have less freedom to choose their walking speed as walking passed the opposing pedestrians becomes difficult \citep{lam2003generalised}. When the flow ratio of both directions is balanced (e.g. close to 0.5), flows from both directions exhibit the highest level of congestion \citep{feliciani2018universal}. When total pedestrian flow is much less than capacity of the walkway, pedestrians have freedom to choose their preferred walking speed. Under free-flow conditions, the effect of bidirectional stream is less significant, but the variation of walking speed remains large \citep{lam2003generalised} due to variations in age, gender, attitudes and physical capabilities \citep{tanaboriboon1991, Lam2002}. As the total pedestrian flow becomes larger, the walkway tends to become more congested and the bidirectional impact becomes more significant. {\color{black} In this study, we assume equal unidirectional capacities on a bidirectional link therefore, it does not capture the link capacity variation based on direction such as walking uphill or downhill.

Several studies in the past have explored pedestrian bidirectional traffic dynamics through mostly microscopic models \citep{lammel2015model,wagoum2015jupedsim, zhu2016study, jin2017simulating, lu2017study,tao2017cellular} and macroscopic models \citep{xu2018simulation,hanseler2013aggregated,hanseler2014macroscopic}. Social force model has been widely adopted as a microscopic model to reproduce bidirectional effects including the look ahead phenomenon \citep{taherifar2019macroscopic}, group behavior \citep{helbing1995social,huang2018social}, and lane formation \citep{hoogendoorn2015continuum,saberi2015,guo2016uni}. \citet{haghani2020empirical1} and \citet{haghani2020empirical2} provide extensive reviews of pedestrian crowd dynamics. The trade-off between the complexity of microscopic behaviour and the computational requirements calls for development of macroscopic models for large-scale pedestrian crowd systems. Cell transmission model maintains some of the spatial and microscopic pedestrian  complexity while requiring extensive computational resources \citep{hanseler2013aggregated,xu2018simulation,han2021pedestrian} especially for large-scale networks \citep{hanseler2014macroscopic}. Like the work presented in this study, \citet{jeanbart2018multi} proposed a macroscopic pedestrian model to optimize pedestrian moving walkways. However, the bidirectional effects were only captured through a reduction in the corridor width.} 

\section{Methodology} \label{sec:method}
Deterministic UE relies on two main assumptions, i.e. travelers have perfect information and behave homogeneously \citep{Sheffi1985}. In pTAP under UE condition, these assumptions are difficult to hold because in reality travelers may not have full knowledge on every possible routes and all travelers are less likely to have identical response to the same traffic condition. Stochastic UE (SUE) relaxes the first assumption by adding randomness into the path travel time \citep{Sheffi1985,connors2007sensitivity,clark2002sensitivity,cascetta2006models}. The second assumption, that expects travelers to behave identically given the same traffic flow condition, is difficult to hold in the pTAP because walking speeds can vary and the probability of lane formation could significantly impact link travel times. A stochastic pVDF can, however, relax the homogeneous behaviour assumption. A typical pedestrian link often accommodates bidirectional flow. Therefore, it is essential to capture the interaction between two opposing streams. In this section, we present four different formulations of the pTAP with: (i) deterministic pVDF with symmetric link interaction, (ii) stochastic pVDF with symmetric link interaction, (iii) deterministic pVDF with asymmetric link interaction, and (iv) stochastic pVDF with asymmetric link interaction.

We first formally define the concept of bidirectional stream. 

\begin{defi}[Bidirectional stream]
A pair of links $a=(i,j) \in A$, from node $i$ to node $j$, $a'=(i',j') \in A$, from node $i'$ to node $j'$ are said to be on the same bidirectional stream if and only if $i=j'$ and $j=i'$. Links which belong to the same bidirectional stream are assumed to have identical link characteristics, i.e. capacity $c_a = c_{a'}$, free-flow travel time $\tau_a=\tau_{a'}$, and length $l_{a} = l_{a'}$.
\end{defi}

{\color{black}In this study, a pedestrian pathway or walkway refers to a dedicated walking track that pedestrians can walk along. A corridor refers to a pathway with physical barrier along both sides of the path. Each bidirectional traffic consists of two directional flows. The reference direction is the direction of interest that moves against the opposite or counter direction on the same bidirectional stream.} Since all pedestrian pathways are assumed to be usable in both directions, all links in the network are expected to belong to a bidirectional stream. Links on the same bidirectional stream are expected to have some interaction, e.g. if one is congested, links on the same bidirectional stream are expected to be congested as well. However the travel times of both links may be or may not be equal depending on the pVDF which will be further discussed in Section \ref{sec:pVDF}. Further, links that belong to the same bidirectional stream share the same walking infrastructure, thus their pVDF attributes are identical.

\subsection{Network representation} \label{sec:networkreprepresentation}
A car traffic network is generally considered as a directed graph $G=(N,A)$ consisting of nodes $N$ and links $A$. Network representation is a widely used building block in TAP \citep{jayakrishnan1994faster,mitradjieva2013stiff} due to its efficiency. Pedestrian traffic has two distinct characteristics from car traffic. Firstly, a car lane is for one way traffic while a pedestrian walkway is for two ways traffic. So a car traffic link accommodates only unidirectional traffic while a pedestrian traffic link has to accommodate bidirectional traffic. Secondly, cars can only traverse on predetermined lane while pedestrians can occasionally form lanes on their own. 

In a city footpath network, urban furniture such as bench, bus stop, or mailbox can disrupt traffic flow which results in a lower effective link capacity than its original link capacity. Traffic signals are implemented by a lower free flow speed on a crossing link to implement waiting time as additional time delay. Each link should have capacity, free-flow travel time, and length attributes. We create an algorithm to convert a road network to a pedestrian footpath network as shown in Figure \ref{fig:footpathgenschematic}. See Appendix~B for further details of the proposed footpath network generation algorithm.

\begin{figure}[pos=htp]
\centering
\includegraphics[width=\textwidth]{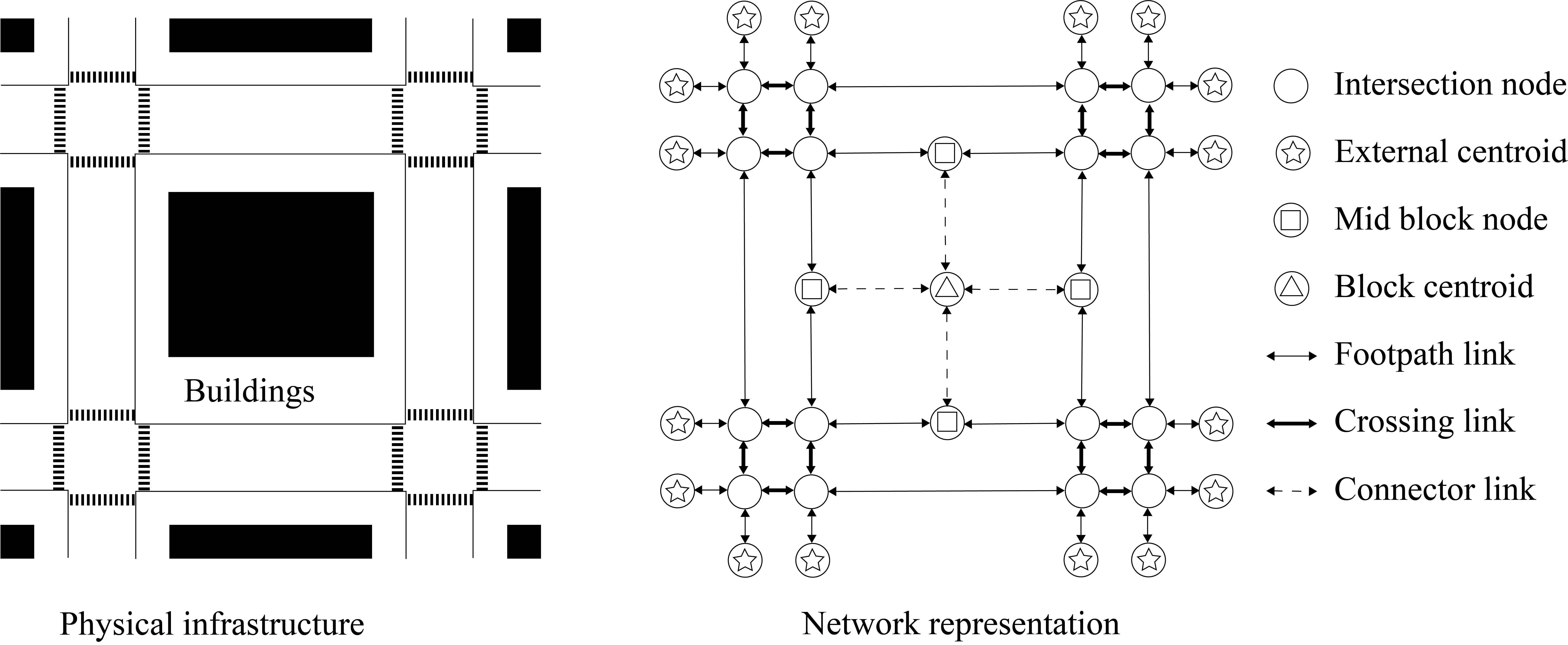}
\caption{Comparison between physical infrastructure and network representation of footpath}
\label{fig:footpathgenschematic}
\end{figure}
External and block centroids denote the beginning or the end of every trips. External centroids are points of interest outside the network. Block centroids represent how many trips generated from/to each area. Intersection and mid block nodes denote intermediate node along all paths. Footpath link denotes physical area between curbs and buildings for foot traffic, crossing link denotes footpath on roads at junctions or intersections, and connector link facilitate connections between physical footpath links and model block centroids. We will further discuss about bidirectional nature of pedestrian traffic in Section \ref{sec:detpVDF} and simultaneous lane formation in Section \ref{sec:stochpVDF}.

\subsection{Pedestrian VDF} \label{sec:pVDF}
The context of pTAP, the link performance function has to incorporate flows from a bidirectional stream with capacity and free-flow travel time attributes. Travel time and flow relationship is established by fitting deterministic pVDFs to empirical data. Both symmetric and asymmetric pVDFs are proposed. We preserve the perfect information assumption and relax an homogeneous behaviour assumption of deterministic UE by proposing stochastic pVDFs as will be further discussed in this section.

\subsubsection{Deterministic pVDF} \label{sec:detpVDF}
We propose two forms of deterministic pVDFs for symmetric and asymmetric link interaction. The basic form of the pVDFs is a strictly monotonic function. {\color{black}Both existence and uniqueness properties of the UE-pTAP solution hold for the symmetric pVDF. However, empirical pedestrian data suggest that pedestrian interactions are better represented with a non-monotonic asymmetric pVDF which is consistent with the previous findings in the literature \citep{Lam2002,kretz2006experimental,helbing2009ped} that will be further discussed in Section 4.1.} Asymmetric pVDF may be more realistic than symmetric pVDF, but it no longer guarantees the existence and uniqueness properties of the UE-pTAP solution. 

We first define symmetry in pVDF and a generic deterministic symmetric pVDF. 

\begin{defi}[Symmetric pVDF]
\label{sympVDF}
Consider a pair of links $a,a' \in A$ which belong to the same bidirectional stream. We say that a pVDF $t_a(x_a,x_{a'})$ is symmetric if and only if 
\begin{equation}\label{eq:vdf_sym}
\hspace{150pt} t_{a}(x_{a},x_{a'}) = t_{a'} (x_{a'},x_{a})
\end{equation}
\end{defi}

\begin{defi}[Deterministic Symmetric pVDF]
Consider a pair of link $a, a' \in A$ which belong to the same bidirectional stream. Let $\tau _{a}$ denote the free flow travel time of link $a$; let $c_{a}$ denote the capacity of  $a$; and let $\alpha$ and $\beta$ be model parameters. We denote 	
\begin{equation}
\hspace{130pt}t_{a}\left(x_{a},x_{a'}\right) = \tau _{a} \left( 1+\alpha \left(\frac{x_{a}+x_{a'}}{c_{a}}\right)^{\beta} \right),
\label{eq:detvdf_sym}
\end{equation}    	
the deterministic symmetric pVDF of link $a \in A$.
\end{defi}

Equation \eqref{eq:vdf_sym} represents the symmetry property of the link interaction in pVDF. Equation \eqref{eq:detvdf_sym} represents a pVDF form with symmetric link interaction that is extended from the well-known BPR function with $\alpha$ representing the ratio of travel time per unit distance at practical capacity to that at free flow while $\beta$ determines how fast the average travel time increases from free-flow to congested conditions. Links from both directions of the same bidirectional stream have identical travel times as required by Equation \eqref{eq:vdf_sym}. Further, Equation \eqref{eq:detvdf_sym} implies that flows from either direction have equivalent impact on travel time of both directions which complies with Definition \ref{sympVDF}. 

Existence and uniqueness of a UE-TAP solution are important properties that have been discussed in many past studies \citep{aashtiani1981equilibria,Sheffi1985, PatrikssonMichael1994Ttap}. \citet{Sheffi1985} proves that the first order condition $\frac{dz(x)}{dx} = 0$ of the Beckmann formulation is equivalent to the UE condition. Therefore, the UE solution exists at the local minimum points. Specifically, to prove the existence of a solution under UE conditions, \citet{Sheffi1985} assumes that travel time on any link depends on the flow of that link only. {\color{black} Consider a pair of links $b, b' \in A$ which belong to the same bidirecctional stream.}
\begin{equation} \label{eq:sheffi_existence}
\hspace{140pt} \frac{\partial z(x)}{\partial x_{b}} = \frac{\partial}{\partial x_{b}} \sum_{a \in A} \int_{0}^{x_{a}} t_{a}( \omega ) d \omega = t_{b}
\end{equation}
However, in the context of symmetric pVDF from Equation \eqref{eq:detvdf_sym}, Equation \eqref{eq:sheffi_existence} is no longer true since the travel time on a link is a function of flow from both reference direction and flow from the counter direction as shown in Equation \eqref{eq:sheffi_disproveexist}. 
\begin{equation} \label{eq:sheffi_disproveexist}
\hspace{150pt} \frac{\partial z(x)}{\partial x_{b}} = t_{b}(x_{b},x_{b'}) + \frac{\tau _{b} \alpha x_{b'}}{c_{b}}
\end{equation}
\citet{aashtiani1981equilibria} proves that at least a solution exists, if a network $G$ is strongly connected and VDFs are non-negative continuous functions. Equation \eqref{eq:detvdf_sym} complies with this definition. So we are unable to prove the solution existence under \citet{Sheffi1985} interpretation, but the symmetric pVDF has at least a solution under \citet{aashtiani1981equilibria} interpretation. 

In order to prove that UE problem has a unique solution, \citet{Sheffi1985} shows that the Beckmann's formulation is strictly convex under two conditions expressed in Equations \eqref{eq:sheffi_uniqueness_a} and \eqref{eq:sheffi_uniqueness_b}.
\begin{subequations}\label{eq:sheffi_uniqueness}
    \begin{align}
        \textrm{\hspace{160pt}} &\frac{\partial t_{a}(x_{a})}{\partial x_{b}} = 0 \textrm{\qquad} \forall a \neq b \label{eq:sheffi_uniqueness_a} \\
        &\frac{\partial t_{a}(x_{a})}{\partial x_{a}} > 0 \textrm{\qquad} \forall a \label{eq:sheffi_uniqueness_b}
    \end{align}
\end{subequations}

The proposed symmetric pVDF complies with Equation \eqref{eq:sheffi_uniqueness_b} since it is a monotonically increasing function. But link travel time depends on flow from both reference and counter directions, so it does not comply with Equation \eqref{eq:sheffi_uniqueness_a}. \citet{PatrikssonMichael1994Ttap} shows that if the link travel time functions are strictly monotonic, then objective function in the Beckmann's formulation is strictly convex. Similarly, \citet{aashtiani1981equilibria} also proves that a UE solution is unique if a network $G$ is strongly connected and VDFs are strictly monotonic functions. Since the proposed symmetric pVDF is both non-negative and strictly monotonic, it implies to have both existence and uniqueness properties. In the next section, we relax the homogeneity behaviour assumption using a stochastic function for a more realistic pVDF.

\subsubsection{Stochastic pVDF} \label{sec:stochpVDF}
The self-organization phenomena is unstable in nature and thus, creates spatio-temporal variations in pedestrian speeds. To incorporate this behaviour within the pTAP, a stochastic travel time function, denoted $T_{a}$ for link $a \in A$, which encapsulates various degrees of lane formation with a standard deviation $\sigma_{a}$ is proposed. The spread of travel time distribution can be expressed with its variance $var(T_{a})$ for any link $a \in A$, and its covariance as $cov(T_{a}, T_{a'})$ for any pair of links $a, a' \in A$. Formally, the variance of link $a$ is defined as
\begin{equation} \label{eq:variance_link}
\hspace{170pt} var(T_{a}) = \sigma_{a}^{2}
\end{equation}
{\color{black}The covariance of link $a$ and link $a'$ can be expressed as} 
\begin{equation} \label{eq:covariance_link} \color{black}
\hspace{100pt} cov(T _{a}, T _{a'}) = \rho _{a,a'} \times \sigma _{a} \times \sigma _{a'}
\end{equation}
\begin{equation} \label{eq:correlation_link} \color{black}
\hspace{100pt} \rho _{a,a'} = 
   \left \{
      \begin{aligned}
        &1 , && \text{if $a$ and $a'$ are on the same bidirectional stream}\\
        &0, && \text{otherwise}
      \end{aligned} \right.
\end{equation}

{\color{black} where $\rho_{a,a'}$ denotes the correlation between link $a$ and link $a'$. To capture the variability of the travel time between travellers, we assume that the correlation only exists between the links on the same bidirectional stream as shown in Equation \ref{eq:correlation_link}. Every pair of links in a bidirectional stream has an identical correlation $\rho = 1$. This assumption may not be true for some specific situations (like ramps or stairways with specific directional characteristics) but for general bidirectional footpaths, this should be sufficiently reasonable. In the future, other correlation structures can be explored, tested, or even verified with empirical data. Under assumptions that link $i$ and link $i+1$ are on the same bidirectional stream where $i \in \{1, 3, 5, ..., A-1\}$, the covariance matrix $\Sigma$ of all links $A$ in the entire network $G$ can be expressed as.}
\begin{equation} \label{eq:covaraincemat} \color{black}
\begin{aligned}
\Sigma = 
    \begin{pmatrix}
        var(T_{1}) & cov(T_{1},T_{2}) & \cdots  &cov(T_{1},T_{|A|}) \\
        cov(T_{2},T_{1}) & var(T_{2}) & \cdots & cov(T_{2},T_{|A|}) \\
        \vdots & \vdots  & \ddots  & \vdots \\
        cov(T_{|A|-1},T_{1}) & cov(T_{|A|-1},T_{2}) & \cdots  & cov(T_{|A|-1},T_{|A|}) \\
        cov(T_{|A|},T_{1}) & cov(T_{|A|},T_{2}) & \cdots  & var(T_{|A|})
    \end{pmatrix} 
     =
    \begin{pmatrix}
        \sigma _{1} ^{2} &  \sigma _{1} \times \sigma _{2} & \cdots   & 0 \\
         \sigma _{2} \times \sigma _{1} & \sigma _{1} ^{2} & \cdots  &  0 \\
        \vdots & \vdots  & \ddots  &  \vdots \\
        0 & 0 & \cdots  &   \sigma _{|A|-1} \times \sigma _{|A|} \\
        0 & 0 & \cdots &\sigma _{|A|} ^{2}
    \end{pmatrix}
\end{aligned}
\end{equation}
The stochastic pVDF is defined as a log-normal distribution consisting of deterministic travel time function as expected value and a constant covariance matrix as shown in Equation \eqref{eq:covaraincemat}. $t_{a}$ establishes an expected travel time value while the covariances matrix regulates the dispersion of travel time.

\begin{equation} \label{eq:stochVDF}
 \hspace{120pt} T_{a}(x_{a},x_{a'}) \sim \mathcal{LN}(t_{a}(x_{a},x_{a'}), \Sigma)  \textrm{\hspace{40pt}} \forall a \in A
 \end{equation}

\begin{prop}[Covariance of links on the same bidirectional stream] 
Consider a pair of links $a, a' \in A$ which belongs to the same bidirectional stream. If the link travel times follow a symmetric pVDF as given by Equation \eqref{eq:vdf_sym}, then 
\begin{equation} \label{eq:covariancebi}
   \hspace{180pt} \sigma _{a} = \sigma _{a'}
\end{equation}
From equations \eqref{eq:variance_link} and \eqref{eq:covariance_link}, if standard deviation of two links are equal, variance and covariance of both links are equal as well. 
\begin{proof}
To prove that two links on the same bidirectional stream have the same covariance, observe that two links on the same bidirectional stream have identical travel time functions as given in Equation \eqref{eq:detvdf_sym}. This implies that both links have equal standard deviation and covariance as well. 
\end{proof} 
\end{prop}

Various probability distributions have been explored in the literature to capture the variability of travel time including Normal, Log-normal \citep{mei1993lognormal,wang2012speed,srinivasan2014finding}, Gamma \citep{mazloumi2010using,kim2015compound}, Weibull \citep{al2006new}, and Burr \citep{susilawati2013distributions}. The log-normal distribution is selected in our proposed pVDF to ensure non-negativity condition while path travel time can still follow an approximated log-normal distribution given the link additivity assumption. This distribution is skewed toward higher values; therefore, the travel time given any flow is equal or greater than the free flow travel time. Each link has a mean travel time derived from a deterministic pVDF and a standard deviation. Although link travel times are assumed to follow a log-normal distribution, the sum of link travel times on a path does not necessarily result in a log-normal distribution for the path travel times. Although a closed-form formulation of the log-normal sum does not exist, it can be approximated by a log-normal distribution. Several studies in the literature propose various approximating methods \citep{fenton1960sum, schwartz1982distribution, beaulieu1995estimating, mehta2007approximating, asmussen2008asymptotics}. Fenton-Wilkinson is a well-known approximation method that enables the estimation of the parameters for a single log-normal distribution as the log-normal sum as shown in Equation \eqref{eq:fentonwilkinson}. {\color{black} We denote $N$ the set of nodes in the network, $q_{rs}$ travel demand from origin node $r \in N$ to destination node $s \in N$, $W = \{(r,s) : r \in N, s \in N, q_{rs} > 0\}$ the set of OD pairs in the network, and $k \in \Pi_{rs}$ the path connecting OD pair $(r,s) \in W$. The approximation results in a mean ($M_{k, rs} $) and a variance ($D_{k, rs} ^{2}$) for the path travel time from the sum of log-normally distributed link travel times. }
\begin{subequations}\label{eq:fentonwilkinson} \color{black}
\begin{align}
\textrm{\hspace{120pt}} & D_{k,rs}^{2} \approx \ln \Bigg ( \frac{\sum_{a \in \hat{A} _{k,rs}} e^{2 \mu _{a}+\sigma _{a}^{2}}(e^{\sigma _{a}^{2} -1})}{(\sum_{a \in \hat{A} _{k,rs}} e^{\mu _{a} + 0.5 \sigma _{a}^{2}})^{2}} \Bigg ) \qquad \forall (r,s) \in W, \forall k \in \Pi_{rs}  \\
& M_{k,rs} \approx \ln \Big ( e^{\mu _{a} + 0.5 \sigma _{a}^{2}} \Big ) - \frac{D_{fw}^{2}}{2} \qquad \forall (r,s) \in W, \forall k \in \Pi_{rs}
\end{align}
\end{subequations}
{\color{black}where $\hat{A} _{k,rs}$ denotes links on a path $k$ connecting OD pair $(r,s)$, $M_{k, rs}$ denotes an approximation of the mean of the path travel times, $D_{k, rs}^{2}$ denotes an approximation of the variance of the path travel times. The Fenton-Wilkinson method provides reasonable accuracy for the PDF in its tail portion, but suffers from inaccuracies in the head portion \citep{beaulieu1995estimating}. The proposed pTAP framework assumes additive link cost property, so the path travel times can be approximated as the sum of the link travel times that are log-normally distributed. Using the Fenton-Wilkinson method, the path travel time is also approximated with a log-normal distribution.}  

\subsection{pTAP with stochastic symmetric pVDF}
In this section, we present an alternative pTAP formulation wherein the pVDF is assumed to be stochastic and symmetric to capture the stochastic nature of self-organization phenomena.

\begin{defi}[Stochastic Symmetric pVDF]
\label{sspVDF}
Consider a pair of link $a, a' \in A$ which belong to the same bidirectional stream. Let $T_{a}$ and $t_{a}$ denote the stochastic and the deterministic link travel time functions of  $a$, $x_{a}$ and $x_{a'}$ denote the link flow of $a$ and $a'$, and $\sigma _{a}$ denotes the standard deviation of $a$. Let  $\phi$, $\gamma$, $\lambda_{t}$ be model parameters. We define the stochastic symmetric pVDF of link $a \in A$ as 	

\begin{equation}
\label{eq:stocvdf_sym}
T_{a} (x_{a},x_{a'}) \sim \mathcal{LN}(t_{a}(x_{a},x_{a'}),\sigma _{a} (x_{a},x_{a'})) \quad \textrm{with} \quad \sigma _{a} (x_{a},x_{a'}) = \tau _{a} \phi \mathrm{e} ^{\gamma \big ( \frac{x_{a} + x_{a'} }{c_{a}} - \lambda _{t} \big ) ^{2}}
\end{equation}
\end{defi}

Observe that in the proposed definition of stochastic symmetric pVDF, the deterministic travel time function $t_a$ represents the expected value of the travel time on link $a \in A$. Further, the standard deviation of the travel time on link $a \in A$, $\sigma_a$ is defined as a symmetric function of link flows of the corresponding bidirectional stream. $\phi$ relates to the magnitude of the standard deviation of the travel time. Decreasing $\phi$ will flatten the curve.{\color{black}$\gamma$ determines the extent of the standard deviation variability. The higher $\gamma$ is, the more concentrated the standard deviation of travel time becomes which means lower variability. $\lambda _{t}$ determines the ratio of flow over capacity that has the maximum standard deviation. If $\lambda _{t}$ increases, the peak of standard deviation will shift to the right. Both $\gamma$ and $\lambda _{t}$ are unitless parameters.}

We next show in Propositions \ref{existss} and \ref{uniquess} that using the proposed stochastic symmetric pVDF, the existence and the uniqueness of the corresponding UE-pTAP solution is guaranteed. The proofs of these propositions are based on the Existence and the Proximity Theorems in \citet{watling2002second}. \citet{watling2002second} focuses on characteristics of cost function with expected travel time, e.g., twice continuously differentiable or monotonically increasing function. In our context, both Propositions \ref{existss} and \ref{uniquess} focus on expected travel time of the stochastic pVDF.

\begin{prop}[Existence condition for pTAP with stochastic symmetric pVDF]
\label{existss} \color{black}
Consider a pTAP formulation with stochastic symmetric UE-pTAP as defined in Definition \ref{sspVDF}. If the pVDF consists of $t_a$, $\forall a \in A$ that are continuous and twice differentiable and all have continuous second derivatives, then at least one UE-pTAP solution exists.
\begin{proof}

    \begin{align*} 
        \hspace{130pt} t_{a}(x_{a},x_{a'}) & = \tau _{a} \Big ( 1+\alpha \Big(\frac{x_{a}+x_{a'}}{c_{a}}\Big)^{\beta} \Big ) \\ 
        \frac{\partial}{\partial x_{a}} t_{a}(x_{a},x_{a'}) &= \frac{\tau _{a} \alpha \beta}{c_{a}} \Big(\frac{x_{a}+x_{a'}}{c_{a}}\Big)^{\beta -1} \\
        \frac{\partial ^{2}}{\partial x_{a} ^{2}} t_{a}(x_{a},x_{a'}) &= \frac{\tau _{a} \alpha \beta (\beta - 1)}{c_{a} ^{2}} \Big(\frac{x_{a}+x_{a'}}{c_{a}}\Big)^{\beta -2} \\
        \because \hspace{20pt} \tau _{a}, \alpha, c_{a} > 0 \hspace{20pt} & \land \hspace{20pt} \beta \geq 1 \hspace{20pt}  \land \hspace{20pt} x_{a}, x_{a'} \geq 0 \\
        \frac{\partial ^{2}}{\partial x_{a} ^{2}} t_{a}(x_{a},x_{a'}) & \geq 0
    \end{align*}
\end{proof}
This shows that $t_{a}$ is continuous and has a non-negative continuous second derivative if parameter $\beta$ is greater or equal to one. The symmetric pVDF ensures satisfy the existence condition as long as it is a polynomial with at least degree one.
\end{prop}

\begin{prop}[Uniqueness condition]
\label{uniquess} \color{black}
Consider a UE-pTAP formulation with stochastic symmetric pVDF as defined in Definition \ref{sspVDF}. If the pVDF consists of $t_{a}$ $\forall a \in A$ is a three times continuously differentiable and monotonically increasing function, and if 
\begin{subequations} \label{eq:unique_conditions}
\begin{align}
    \hspace{150pt} \frac{\partial ^{2}}{\partial x_{a} ^{2}} t_{a} & \geq 0 \hspace{20pt} \forall \beta \in \{ (-\infty,0], [1,\infty] \} \label{eq:unique_conditions_a}\\
    \frac{\partial ^{3}}{\partial x_{a} ^{3}} t_{a} & \geq 0 \hspace{20pt} \forall \beta \in \{ [0,1], [2,\infty) \} \label{eq:unique_conditions_b}
\end{align}
\end{subequations}
Then if there exists a UE-pTAP solution, this solution is also unique.
\begin{proof}
    From the proof of Proposition \ref{existss}, the condition \eqref{eq:unique_conditions_a} is already satisfied. We can extend from the second derivative to the third derivative as follows
    \begin{align*}
        \hspace{130pt} \frac{\partial ^{3}}{\partial x_{a} ^{3}} t_{a}(x_{a},x_{a'}) &= \frac{\tau _{a} \alpha \beta (\beta - 1) (\beta - 2)}{c_{a} ^{3}} \Big(\frac{x_{a}+x_{a'}}{c_{a}}\Big)^{\beta -3} \\
        \because \hspace{20pt} \tau _{a}, \alpha, c_{a} > 0 \hspace{20pt} & \land \hspace{20pt} \beta \geq 2  \hspace{20pt} \land \hspace{20pt} x_{a}, x_{a'} \geq 0 \\
        \frac{\partial ^{3}}{\partial x_{a} ^{3}} t_{a}(x_{a},x_{a'}) & \geq 0
    \end{align*}
    This shows that $t_{a}$ is three times continuously differentiable and satisfies condition \eqref{eq:unique_conditions_b} if parameter $\beta$ is greater or equal to two. The symmetric pVDF ensures to satisfy the uniqueness condition as long as it is a polynomial with at least a second degree. $t_{a}$ is also a monotonically increasing function as discussed in Section \ref{sec:detpVDF}
\end{proof}
\end{prop}

The solution of the pTAP under UE with proposed stochastic symmetric pVDF is proven to have both the existence and uniqueness properties. Both Propositions  \ref{existss} and \ref{uniquess} demonstrate that $t_{a}$ ($\forall a \in A$) satisfies several conditions similar to the proof in \citet{watling2002second}. The next subsection explores the asymmetric pVDF, which is more realistic but does not always guarantee existence and uniqueness of the UE-pTAP solution.

\subsection{pTAP with deterministic asymmetric pVDF}
 Here, we introduce a pVDF with asymmetric link interaction by extending Equation \eqref{eq:detvdf_sym} to capture an asymmetric bidirectional impact on link travel times. Asymmetric pVDF has two key distinct characteristics from the symmetric pVDF. Asymmetric pVDF negates an assumption in Definition \ref{sympVDF} in which flows from the reference and counter directions may have different impact on the reference direction  \citep{lam2003generalised}. The asymmetric pVDF exhibits a local peak in travel time at a certain flow range, while this crest is not present in the symmetric pVDF.

\begin{defi}[Deterministic Asymmetric pVDF]
\label{dapVDF}
Consider a pair of link $a, a' \in A$ which belongs to the same bidirectional stream. Let $\tau _{a}$ denotes the free-flow travel time of a link; $c_{a}$ denotes the capacity of link $a \in A$. Let $\alpha, \beta, \mu, \eta _{r}, \lambda _{r}, \eta _{c},$ and $ \lambda _{c}$ be model parameters. We denote

\begin{equation} \label{eq:detvdf_asym} \color{black}
\hspace{80pt} t_{a}(x_{a},x_{a'}) = \tau _{a} \Big ( 1+\alpha \Big(\frac{x_{a}+x_{a'}}{c_{a}}\Big)^{\beta}  + \mu \mathrm{e} ^{\eta _{r} \big ( \frac{x_{a}}{c_{a}}- \lambda _{r} \big ) ^{2} + \eta _{c} \big ( \frac{x_{a'}}{c_{a}}- \lambda _{c} \big ) ^{2}} \Big )
\end{equation}

the deterministic asymmetric pVDF of link $a \in A$
\end{defi}
The first term in Equation \eqref{eq:detvdf_asym} is the same as that in Equation \eqref{eq:detvdf_sym}. The second term shows that as pedestrian flows get closer to a certain value ($\eta _{r}$ for the reference direction and $\eta _{c}$ for the counter direction), the travel time substantially increases. This term is referred to as the bidirectional term. The bidirectional term becomes negligible when flows are further away from these values ($\eta _{r}$ and $\eta _{c}$). The term consists of five parameters: $\mu$ determines the magnitude of the bidirectional impact for the balanced flows. $\eta _{r}$ and $\eta _{c}$ determine the range of flows that exhibit bidirectional impact by regulating the width of the bell-shaped curve base. $\lambda _{r}$ and $\lambda _{c}$ determine the flow ratios that have the highest congestion level in the stream. 

We next extend Definition \ref{dapVDF} to capture the stochastic nature of lane formation.

\subsection{pTAP with stochastic asymmetric pVDF}
Here, we present an alternative pVDF formulation with asymmetric link interaction effects. This motivates the definition of stochastic asymmetric pVDF using a log-normal distribution similar to what was defined in Definition \ref{sspVDF}. 

\begin{defi}[Stochastic Asymmetric pVDF]
Consider a pair of link $a, a' \in A$ which belongs to the same bidirectional stream. Let $T_{a}$ denotes the stochastic link travel times of link $a$, $t_{a}$ denotes the deterministic travel time following asymmetric pVDF from Equation \eqref{eq:detvdf_asym} , $x_{a}$ and $x_{a'}$ denotes link flow of link $a$ and link $a'$, $\sigma _{a}$ denote the link flow standard deviation of link $a$ as defined in Equation \eqref{eq:stocvdf_sym}. Let $\phi, \gamma, \lambda _{t}$ be model parameters. We denote 	
\begin{equation}\label{eq:stocvdf_asym}
     \hspace{120pt} T_{a} (x_{a},x_{a'}) \sim \mathcal{LN}(t_{a}(x_{a},x_{a'}),\sigma _{a} (x_{a},x_{a'})) 
\end{equation}
the stochastic asymmetric pVDF of link $a \in A$.
\end{defi}

Similar to the discussion in section \ref{sec:detpVDF}, UE-TAP with asymmetric pVDF carries the solution existence property since the pVDF is a non-negative continuous function \citep{aashtiani1981equilibria}. But the uniqueness of the solution may no longer hold true since the pVDF is not a strictly monotonic function with respect to flow.

\section{Calibration} \label{sec:calibration}
We use empirical data from a set of controlled bidirectional experiments \citep{zhang2012ordering} to calibrate the proposed pVDFs. The experiments were conducted in 2009 with up to 350 participants. The average free flow speed across the experiments was 1.55$\pm$0.18 m/s. {\color{black} Before each experiment, participants waited in a waiting area on both sides of a 8m corridor. In some of the experiments, participants were instructed to exit the corridor either to the left or right. In other experiments, participants were free to choose any exits (right or left). At the beginning of each experiment, all participants exit the waiting area through a 4m long pathway leading to the corridor to minimize the effect of entrance. When participants reach the end of corridor, they must either turn left or right before returning to the waiting area for the next experiment. The waiting area exit widths, corridor widths, and the number of participants were varied for each run.} Individual pedestrian trajectories were automatically extracted from video footages \citep{boltes2013}. Density and flow measurements were extracted from the trajectory data using area-wide Edie's definitions \citep{Saberi2014}. {\color{black} For more details on the experiments and data, please refer to \cite{zhang2012ordering}.}

\begin{figure}[pos=htp]
    \centering
    \subfloat[b][]{
        \includegraphics[height=5cm]{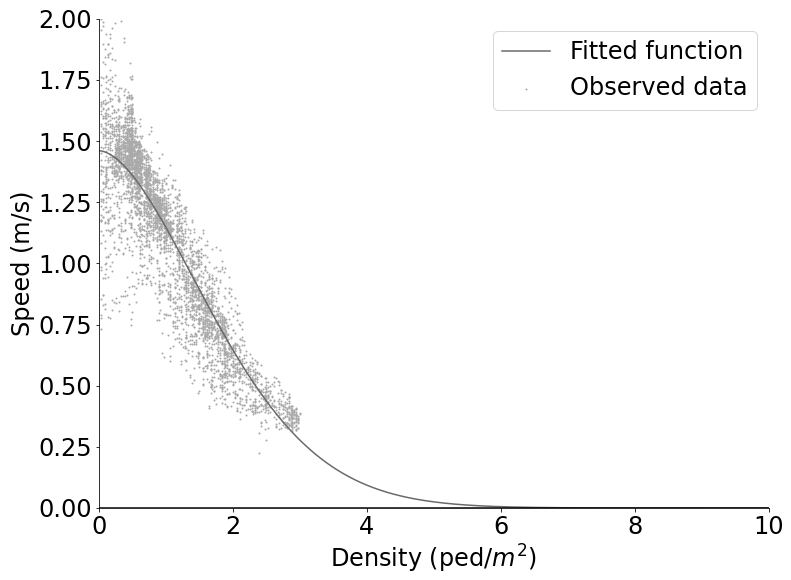}
        \label{subfig:oversaturated1}} 
    \quad 
    \subfloat[b][]{
    	\includegraphics[height=5cm]{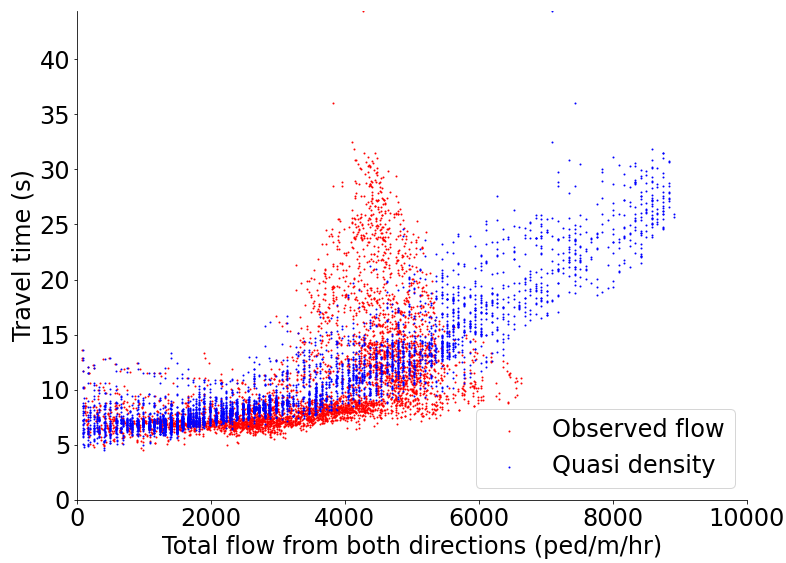}
    	\label{subfig:oversaturated2}} 
    
    \caption{An overview of the empirical data (a) Speed and flow relationship from Equation \eqref{eq:tregenza} (b) Travel time and flow comparison between observed flow and quasi-density}
    \label{fig:ovsersaturated}
\end{figure}

 Here we adopt the approach proposed by \citep{wu2021characterization, kucharski2017estimating} to address the issue with the oversaturated states when calibrating the pVDFs. We translate the observable saturated flows into unobservable oversaturated flows. In pedestrian traffic, the speed and density have a renowned fundamental relationship \citep{cao2017fundamental,nikolic2016probabilistic,Zhang2013,zhang2011transitions,Wong2010}. We utilize the relationship proposed by \citet{tregenza1976design} suggesting that speed has an exponentially decreasing relationship with density as shown in Equation \eqref{eq:tregenza} with three parameters of free flow speed $u_{f}$, $\gamma$, and $\theta$. The speed function can be expressed as
\begin{equation}\label{eq:tregenza}
     \hspace{120pt} u = u_{f} exp \Big( - \Big ( \frac{k} {\theta} \Big ) ^{\gamma} \Big )  
\end{equation}
where $u_{f}$ denotes free flow speed and $\theta$ and $\gamma$ are parameters to be calibrated. The fitted speed function enables derivation of the critical density and flow capacity from Equation \eqref{eq:kc_tregenza} and Equation \eqref{eq:c_tregenza}, respectively. The fitted speed and density relationship shows a reasonably good fit to the empirical data as shown in Figure \ref{subfig:oversaturated1}. The critical density can therefore, be expressed as
\begin{equation}\label{eq:kc_tregenza}
    \hspace{150pt} k_{c} = \frac{\theta}{ \sqrt[\gamma]{\gamma} }
\end{equation}

We also determine the capacity by substituting Equation \eqref{eq:kc_tregenza} and Equation \eqref{eq:tregenza} into the traffic flow fundamental identity $x = u k$.
\begin{equation}\label{eq:c_tregenza}
     \hspace{150pt} c = u_{f} exp \Big ( - \frac{1}{\gamma} \Big ) \Big ( \frac{\theta}{\sqrt[\gamma]{\gamma}} \Big )
\end{equation}

{\color{black}Here, both flow and density increase as demand grows. A surrogate measure known as quasi-density can be used to substitute flow in the pVDF. See \citep{kucharski2017estimating} for more details.} Flow is a multiplication of capacity and ratio of density to critical density as shown in Equation \eqref{eq:kucharski}. 
\begin{equation}\label{eq:kucharski}
     \hspace{150pt} \hat{x}_{i} = c \frac{ k_{i} }{k_{c}}
\end{equation}
Using quasi-density instead of observed flow creates a consistency between the BPR type VDFs and the traffic flow fundamental diagrams as suggested in \citet{wu2021characterization}. From Figure \ref{subfig:oversaturated2}, travel time tends to rise as quasi-density increases while travel time as a function of the observed flow seems to be more dispersed. Travel time from the observed flow measurements increases during the unsaturated flow from 0 - 3,000 ped/m/hr and sharply drop when flow is saturated between 5,000 - 6,000 ped/m/hr. As expected, we are not able to observe the oversaturated state directly from the observed flow as flow is less than 6,000 ped/m/hr while quasi-density estimation suggests oversaturation. Here, we use the concept of quasi-density to calibrate the proposed pVDFs. {\color{black} We apply a least squared optimization method to fit each proposed function to the empirical data. Coefficient of determination ($R^{2}$) and root mean squared error (RMSE) are calculated to evaluate the goodness of fit for each calibrated pVDF. \citet{hoogendoorn2018macroscopic, lam2003generalised, Floetteroed2015} used $R^{2}$ to determine the goodness of fit of their models for estimation of pedestrian flow and density. Both $R^{2}$ and RMSE have also been widely used to determine the goodness of fit of BPR-type VDFs to road traffic data \citep{apronti2015wyoming,so2016estimating,wong2016network}.}

\FloatBarrier

We use the RMSE to compare the difference between the empirical travel time and the predicted travel time by the proposed pVDFs. 
\begin{equation}\label{eq:rmse}
     \hspace{150pt} RMSE = \sqrt{ \sum _{i \in n} ( t_{i} (x _{i},x' _{i}) - \hat{t} _{i})^{2} } 
\end{equation}
where $t_{i} (x _{i},x' _{i})$ denotes the predicted travel time with regard to the flow of the reference direction $x _{i}$ and flow of the counter direction $x' _{i}$ of data point $i$, $\hat{t} _{i}$ denotes the observed travel time of data point $i$, and $n$ denotes the set of observed data. We also use the ($R^{2}$) as another popular goodness of fit measure for evaluation of the model predictions. There are many interpretations of $R^{2}$ \citep{kvaalseth1985cautionary}. Here we use the nonlinear estimation of $R^{2}$ \citep{magee1990r}.
\begin{equation}\label{eq:R2}
     \hspace{150pt} R^{2} = 1 -  \sum _{i \in n}  \frac{( t _{i} (x _{i},x' _{i}) - \hat{t} _{i})^{2}} {( t _{i}(x _{i},x' _{i}) - \bar{t} _{i})^{2}}
\end{equation}
where $\bar{t}$ denotes mean of the observed travel times.

\subsection{Deterministic pVDF}
\begin{figure}[pos=htp]
\centering
\includegraphics[scale=0.36]{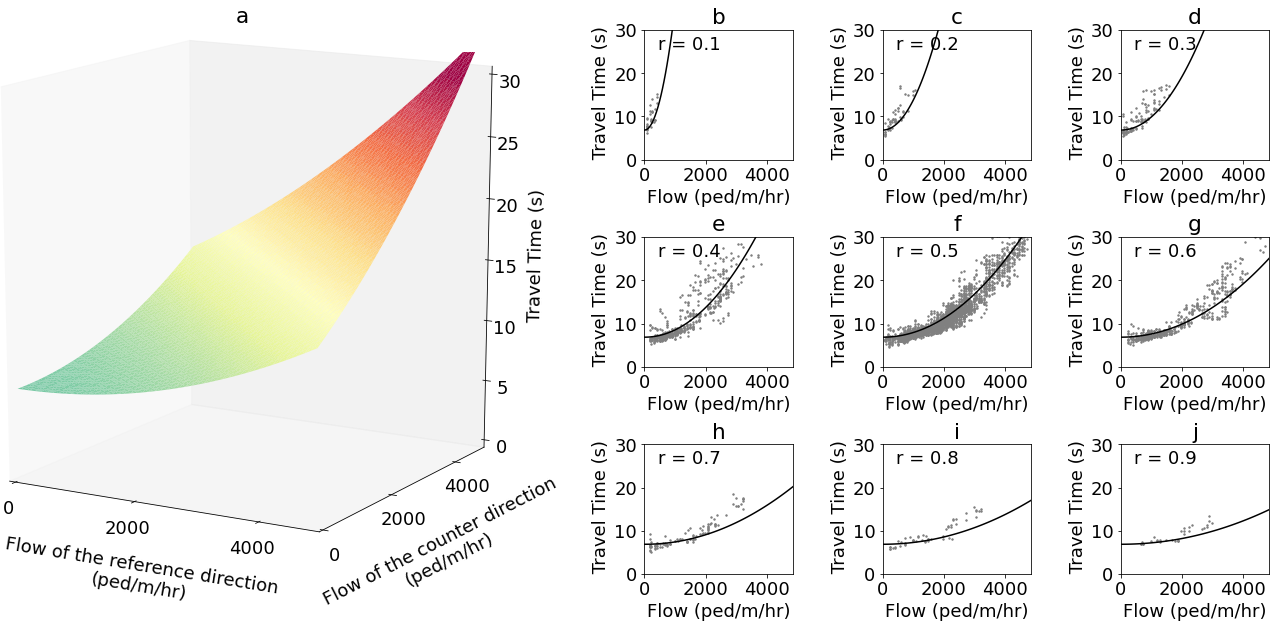}
\caption{Calibrated deterministic symmetric pVDF: (a) peredicted link travel time as a function of flow of the reference and counter directions and (b-j) estimated link travel time as a function of flow of the reference direction for different flow ratios $r$. Measurements are based on a 10m long and 4m wide corridor.}
\label{fig:deterministic_vdf_sym}
\end{figure}

\begin{figure}[pos=htp]
\centering
\includegraphics[scale=0.36]{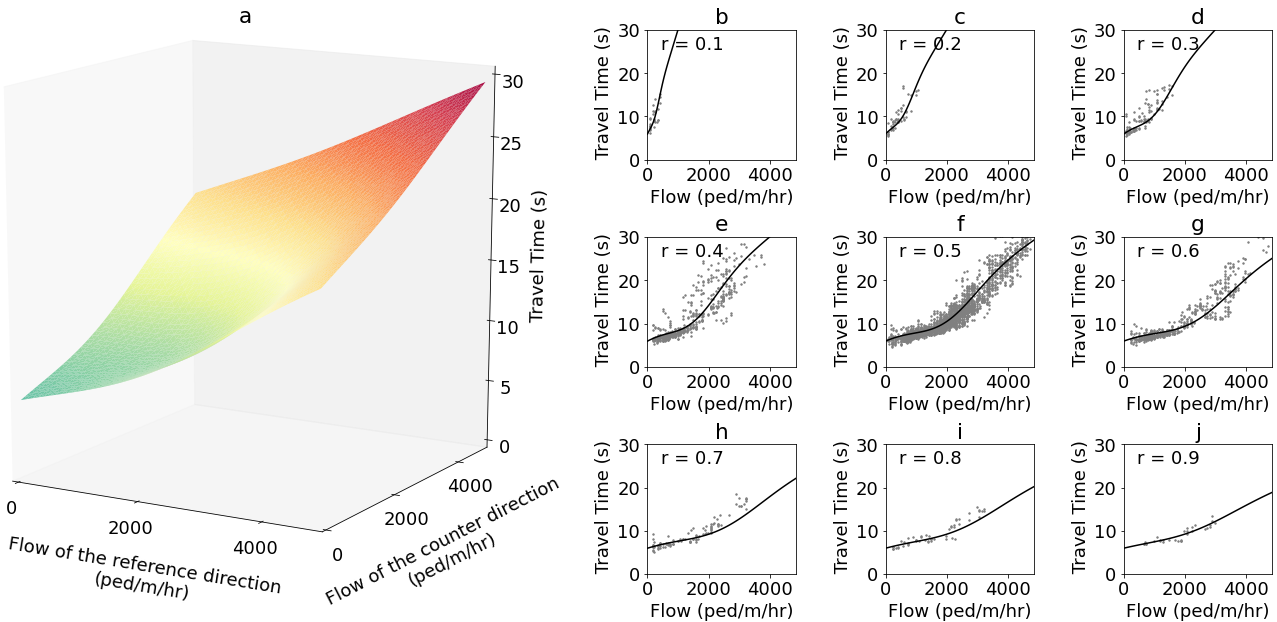}
\caption{Calibrated deterministic asymmetric pVDF: (a) estimated link travel time as a function of flow of the reference and counter directions and (b-j) estimated link travel time as a function of flow of the reference direction for different flow ratios $r$. Measurements are based on a 10m long and 4m wide corridor.}
\label{fig:deterministic_vdf_asym}
\end{figure}

We calculate time-dependent average travel times across several experiments and attempt to fit the proposed symmetric and asymmetric pVDFs to minimize RMSE. The corridor capacity ($c_{a}$) is 4,847 pedestrians/m/hr and free flow travel time  ($\tau _{a}$) is 0.685 seconds.  The calibrated deterministic symmetric pVDF expressed in equation \eqref{eq:detvdf_sym} has $\alpha = 0.949$ and $\beta = 2.031$  as shown in Table \ref{tab:detpVDF} and Figure \ref{fig:deterministic_vdf_sym}. The calibrated deterministic asymmetric pVDF expressed in Equation \eqref{eq:detvdf_asym} is shown in Table \ref{tab:detpVDF} and Figure \ref{fig:deterministic_vdf_asym}. The proposed symmetric pVDF is a monotonically increasing function while the asymmetric pVDF is not monotonically increasing. The proposed asymmetric pVDF consists of two terms. {\color{black} The first term represents the symmetric dynamics and is strictly monotonic, but the second term representing the bidirectional dynamics is not monotonic as shown in Figure \ref{fig:deterministic_vdf_asym}. Based on the calibration results, the proposed asymmetric pVDF generally has monotonic behavior. However, as the ratio of flow over capacity of the reference direction $(x_a/c_a)$ becomes closer to $\lambda_r$  and the flow ratio of the counter direction $(x_a'/c_a)$ becomes closer to $\lambda_c$, the second term becomes negative, hence violating monotonicity assumption as shown in Figure \ref{fig:deterministic_vdf_asym_breakdown}. In this example, the proposed asymmetric pVDF seems to be a weakly monotonically increasing function. However, monotonicity property does not persist in general for the asymmetric pVDF. So, the proposed asymmetric pVDF is non-monotonic.}

\begin{table}[ht]
    \centering 
    \begin{tabular}{c c c c c c c c c c}
    \hline
            pVDF type & $\alpha$  & $\beta$ & $\mu$ & $\eta _{r}$ & $\eta _{c}$ & $\lambda {r}$ & $\lambda _{c}$ & RMSE & $R^{2}$\\ \hline
            Symmetric   & 0.949  & 2.031 & - & - & - & - & - & 0.188 & 0.873            \\ 
            Asymmetric   & 1.658  & 0.997 & -0.836 & -5.447 & -5.737 & 0.415 & 0.394 & 0.422 & 0.885            \\\hline
    \end{tabular}
    \caption{Calibrated results for the deterministic pVDF}
    \label{tab:detpVDF}
\end{table}


\FloatBarrier

\subsection{Stochastic pVDF}
\begin{figure}[pos=htp]
    \centering
    \subfloat[][]{
        \includegraphics[height = 5cm]{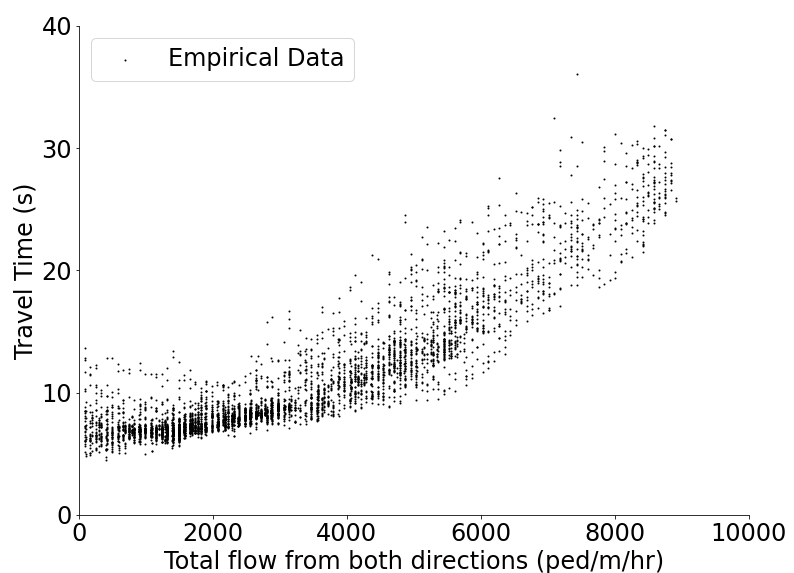}
        \label{subfig:std_sym_a}} 
    \quad 
    \subfloat[][]{
    	\includegraphics[height=5cm]{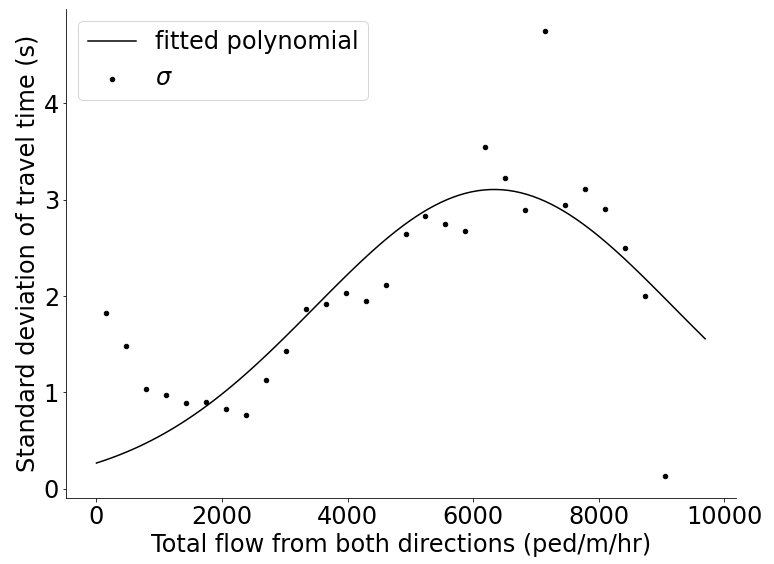}
    	\label{subfig:std_sym_b}} 
    
    \caption{Calibration of the stochastic pVDF. (a) Observed empirical data. (b) Relationship between travel time standard deviation and mean link flow.}
    \label{fig:std_sym}
\end{figure}

Empirical data suggests that for any given pedestrian flow, travel time varies significantly due to the stochastic nature of pedestrian movements in a bidrectional stream. See Figure \ref{subfig:std_sym_a}. Here, we use the estimated travel times from the deterministic symmetric and asymmetric pVDFs as the expected values of log-normally distributed travel times. Thus, at different flows, the travel time distribution has a different expected value and a standard deviation $\sigma$ that is a function of the mean pedestrian total flow. The total flow is defined as the sum of quasi-density from the reference and the counter directions. See Figure \ref{subfig:std_sym_b}.

\begin{figure}[pos=htp]
\centering
\includegraphics[scale=0.4]{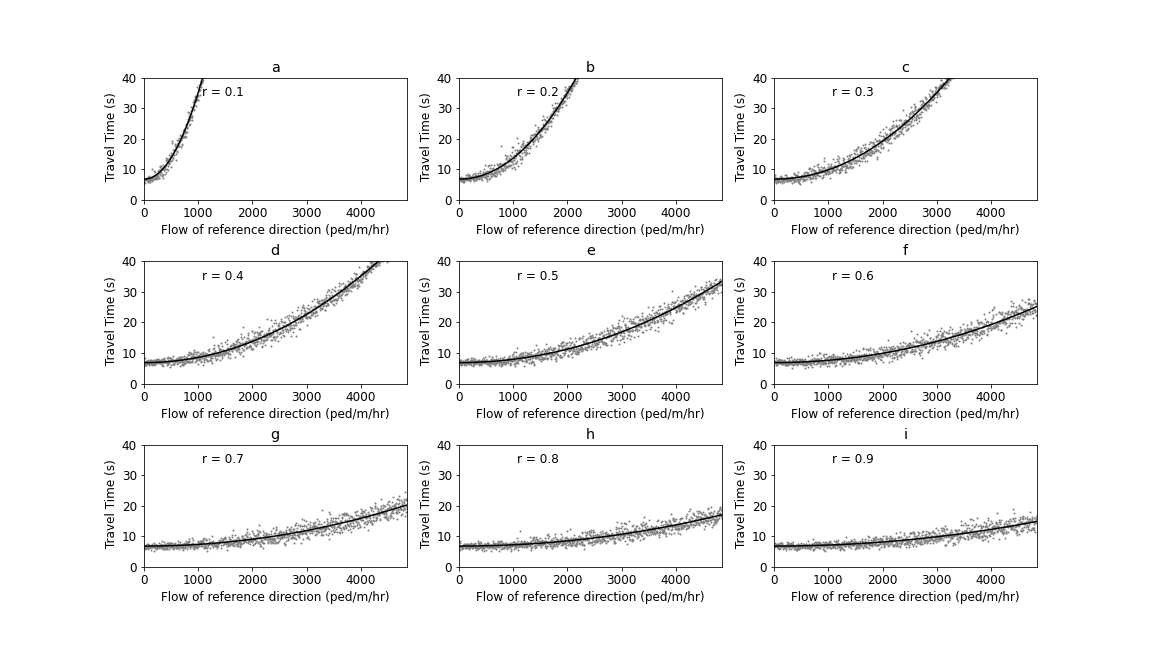}
\caption {Calibrated stochastic symmetric pVDF. Solid line represents the deterministic pVDF. Grey points represent the stochastic sample points drawn from the travel time log normal distribution.}
\label{fig:stochastic_vdf_sym}
\end{figure}

\begin{figure}[pos=htp]
\centering
\includegraphics[scale=0.4]{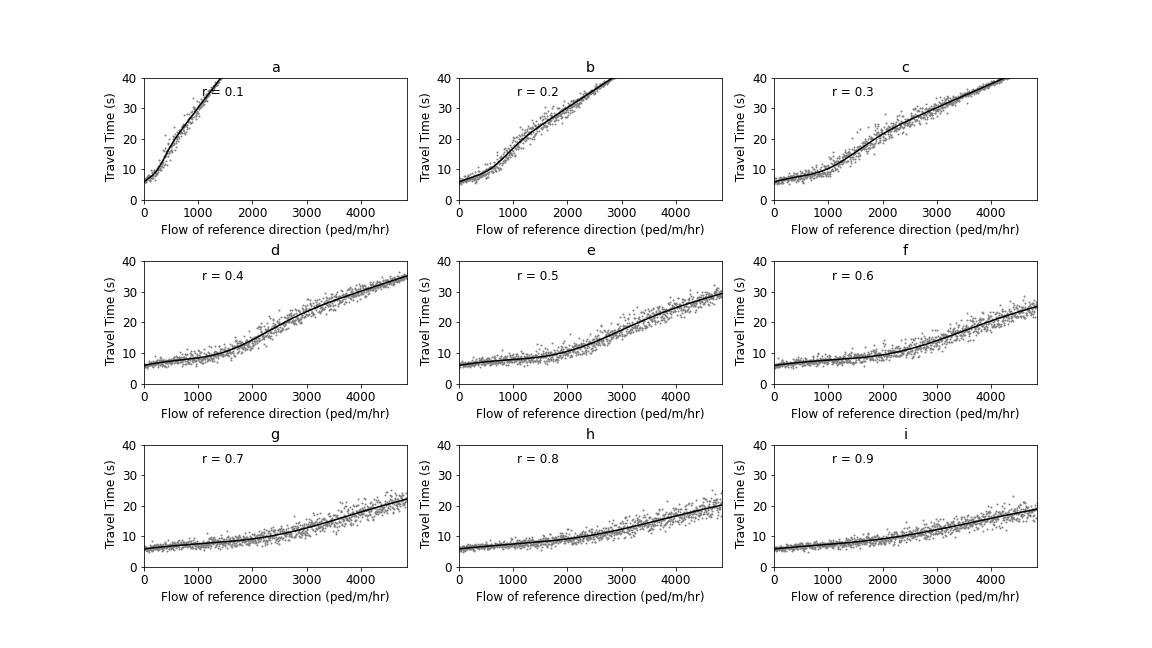}
\caption {Calibrated stochastic asymmetric pVDF. Solid line represents the deterministic pVDF. Grey points represent stochastic sample points drawn from the travel time log normal distribution.}
\label{fig:stochastic_vdf_asym}
\end{figure}

Travel times are likely to be more fluctuating when the total flow is between 5,000 - 7,000 ped/m/hr as shown in Figure \subref*{subfig:std_sym_a}. The standard deviation of travel time also clearly reflects this fluctuation when the total flow is between 5,000 - 7,000 ped/m/hr as shown in Figure \subref*{subfig:std_sym_b}. The calibrated parameters of the standard deviation function expressed in Equation \eqref{eq:stocvdf_sym} are $\phi = 0.454$, $\gamma = 1.439$, and $\lambda_{t} = 1.307$. The stochastic pVDFs can be produced using the calibrated deterministic pVDF and the standard deviation function as shown in Figure \ref{fig:stochastic_vdf_sym} for the symmetric pVDF and Figure \ref{fig:stochastic_vdf_asym} for the asymmetric pVDF.

The standard deviation of travel time is peaked when the total flow from, measured with the described quasi-density approach, is about 6,500 ped/m/hr as shown in Figure \subref*{subfig:std_sym_b}. Both Figure \ref{fig:stochastic_vdf_sym} and Figure \ref{fig:stochastic_vdf_asym} show that when the flow of the reference direction reaches a critical range, travel time becomes widely scattered. For the flow ratio of 0.3, the travel time of the reference direction exhibits the highest variability when the reference direction flow is at 1,900 ped/m/hr. However, for the flow ratio of 0.6, the travel time of the reference direction shows the highest variability when the reference direction flow is at 3,800 ped/m/hr. This clearly demonstrates the significant impact of the flow ratio on bidirectional pedestrian travel times. 

\section{Numerical case studies} \label{sec:case}
\subsection{A small toy network}
\begin{figure}[pos=htp]
    \centering
        {
        \subfloat[][]{
            \includegraphics[scale=0.2]{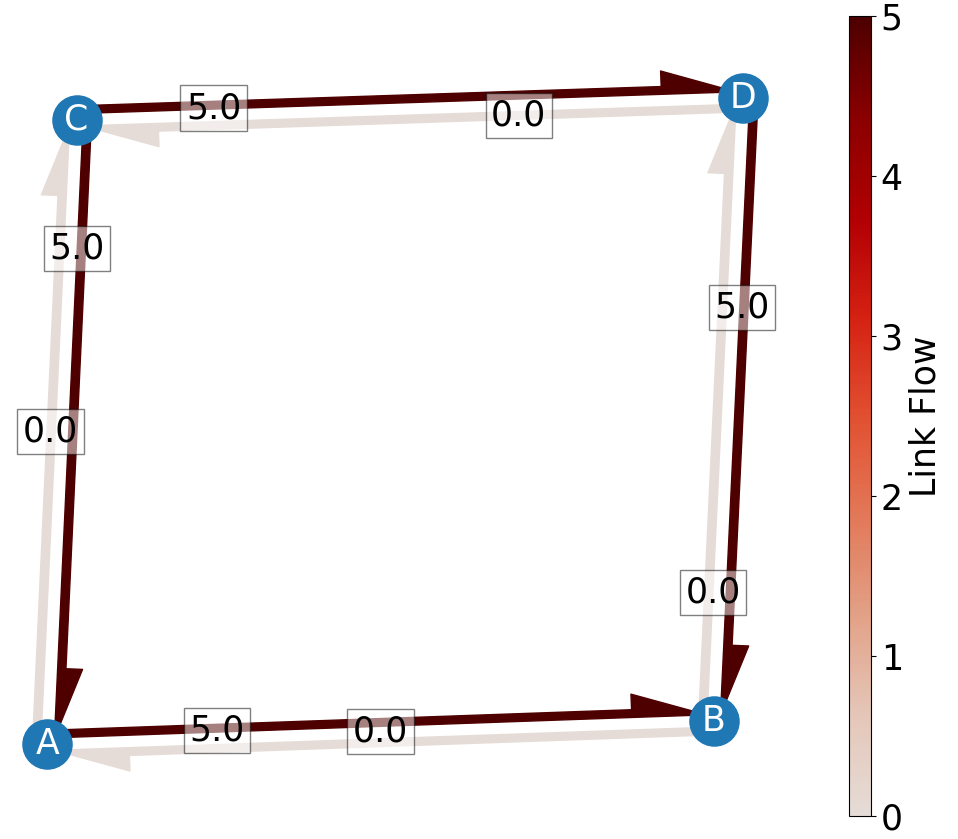}
            \label{subfig:d0}} 
        \hspace{5pt}
        \subfloat[][]{
    		\includegraphics[scale=0.2]{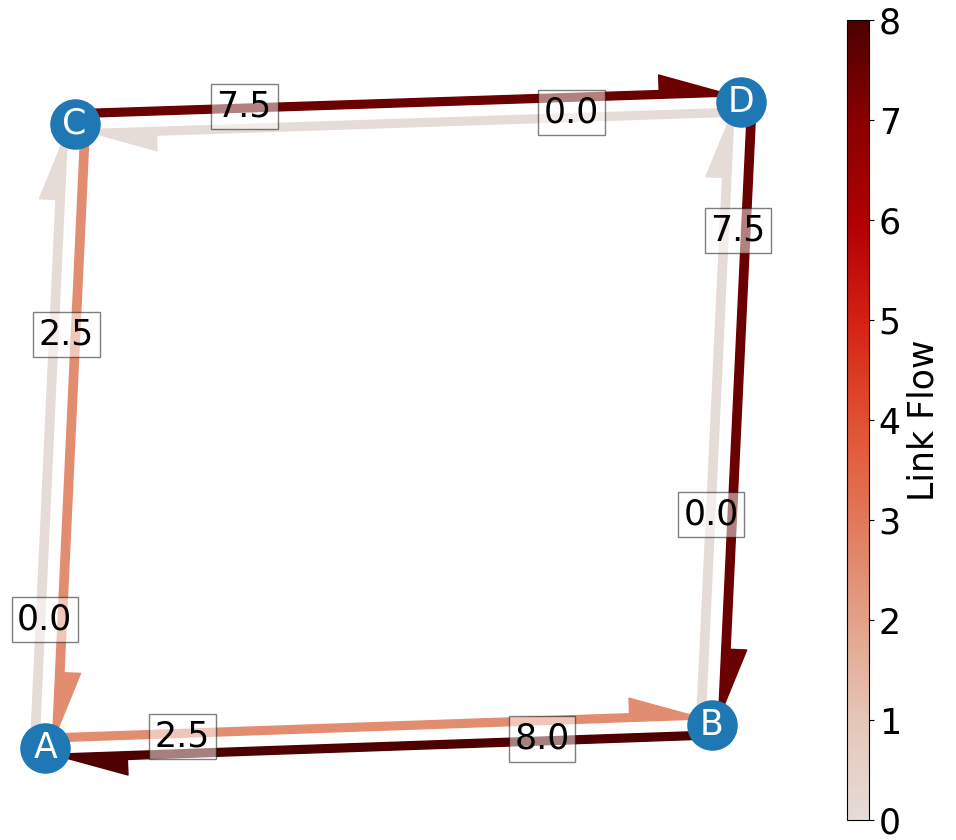}
    		\label{subfig:d1}}
		 \hspace{5pt}
        \subfloat[][]{
    		\includegraphics[scale=0.2]{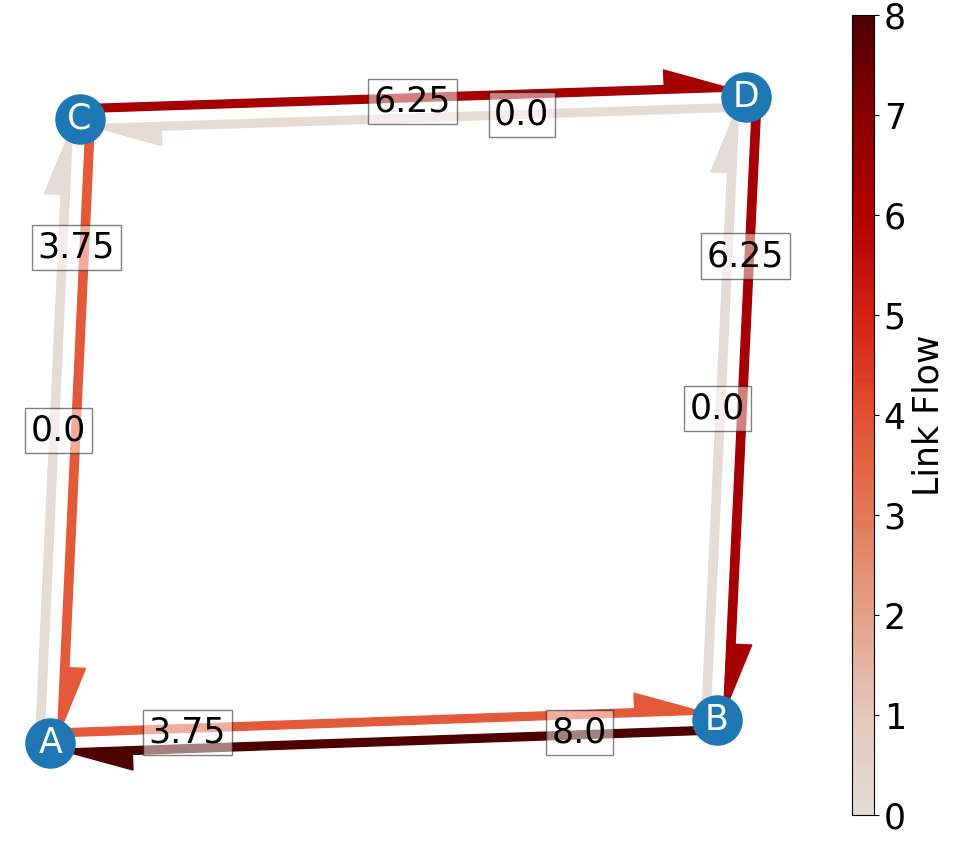}
    		\label{subfig:d1_asym}}
		}
	\caption {Small network application. (a) Case 1: 10 pedestrians travel from node C to B only. The result satisfies UE condition using deterministic symmetric pVDF (b) Case 2: 10 pedestrians travel from node C to B and 8 pedestrians travel from node B to A in the opposite direction. The result satisfies UE condition using deterministic symmetric pVDF (c) Case 3: 10 pedestrians travel from node C to B and 8 pedestrians travel from node B to A in the opposite direction. The result satisfies UE condition using deterministic asymmetric pVDF. black circles represent nodes. Lines with arrow heads represent directional links. Link flows are shown using a color gradient and labels.}
    \label{fig:toy1}
\end{figure}

To demonstrate the impact of bidirectionality on pedestrian route choice, we apply the proposed UE-pTAP framework with deterministic symmetric pVDF to a small network consisting of 4 nodes and 8 links (4 pairs of bidirectional links). Each link in the network is 12m long and 1m wide. We assume a free-flow travel speed of 1.46 $m/s$ and total time of 60 seconds. The network also includes two OD pairs from node C to node B and from node B to node A. We assume a fixed demand of 10 pedestrians going from node C to B. To demonstrate the impact of bidirectional flow, we increase the demand from node B to A from 0 (case 1) to 8 (case 2) pedestrians as shown in Figure \ref{fig:toy1}.

 We also apply the proposed deterministic asymmetric pVDF here. If we assume only a fixed demand of 10 pedestrians travel from node C to B, the flow pattern will split in half similar to case 1. In case 3, we assume a demand of 10 pedestrians travel from node C to B and a demand of 8 pedestrians travel from node B to A. Here, pedestrians will split slightly differently than case 2 due to the asymmetric link interactions and the different form of the pVDF. 

\begin{table}[ht]
    \centering
    \begin{tabular}{c O O O O O O}
    \hline
         & \multicolumn{2}{c }{Case 1} & \multicolumn{2}{ c}{Case 2}   & \multicolumn{2}{ c}{Case 3}  \\ \hline
            link & volume (ped)  & travel time (s) & volume (ped) & travel time (s) & volume (ped) & travel time (s)\\ \hline \\
            A-B  & 5             & 8.47       & 2.5            & 9.37    &    3.75 & 9.87    \\ 
            B-A  & 0             & 8.47        & 8            & 9.37   & 8 & 9.79        \\ 
            C-A  & 5             & 8.47        & 2.5            & 8.28   & 3.75 & 8.26         \\ 
            A-C  & 0             & 8.47        & 0            & 8.28   & 0 & 8.27         \\
            D-B  & 5             & 8.47        & 7.5            & 8.80   &  6.25 & 9.05       \\ 
            B-D  & 0             & 8.47        & 0            & 8.80   & 0 & 9.08         \\  
            D-C  & 0             & 8.47        & 0            & 8.80    & 0 & 9.08        \\ 
            C-D  & 5             & 8.47        & 7.5            & 8.80   & 6.25 & 9.05         \\ \hline
    \end{tabular}
    \caption {Estimated link volumes and travel times across the network for UE-pTAP with either deterministic symmetric pVDF for case 1 and case 2 or deterministic asymmetric pVDF for case 3.}
    \label{tab:num1_table}
\end{table}

\FloatBarrier 
Case 1 represents the control case where opposite link flows do not exist. See Figure \ref{fig:toy1}. Therefore, demand from node C to node B naturally splits in half taking path C-D-B and C-A-B. However, in case 2 where an opposing flow exists on link B-A, path C-D-B experiences higher volume compared to path C-A-B. In Table \ref{tab:num1_table}, we explore link travel time and volume in more details. For case 1, there are two possible paths connecting node C to B. Links C-A and A-B are on one path and links C-D and D-B are on another path. All four links on both paths have the same volume and same travel time. Links on the opposite directions (links D-C, B-D, A-C, B-A) also have the same travel time as well since we use the deterministic symmetric pVDF. Travel time of path C-A-B and path C-D-B are 16.94 seconds in case 1, but path travel times increase to 17.65 seconds in case 2. More pedestrians prefer to take path C-A-B rather than C-D-B to avoid congestion on link A-B. Travel time of link C-A is 8.47 seconds in case 1, but it reduces to 8.28 seconds in case 2. However, travel time of link A-B is 8.47 seconds in case 1, but it increases to 9.37 seconds in case 2. In case 3, not as many pedestrians prefer to take path C-A-B against C-D-B compared to case 2. Unlike case 2, travel time of links on the same bidirectional stream are not the same. For example in case 2, travel time of link A-B and B-A are 9.37 seconds, but in case 3 travel time of link A-B and B-A are 9.87 and 9.79 seconds, respectively. Since link A-B (3.75 pedestrians) is a minor flow while link B-A (8 pedestrians) is a major flow, this implies that congestion affects link A-B more than link B-A. This example demonstrates that by adding additional demand to the opposite direction, all path travel times increase while link travel times may vary differently depending on the flow ratio and the used pVDF. This example clearly shows that increasing volume on link B-A affects the volume on link A-B, demonstrating the impact of bidirectionality.

{\color{black}To quantitatively investigate the influence of demand on the resulting flow patterns, we perform a sensitivity analysis with the deterministic symmetric and asymmetric pVDFs as shown in Table \ref{tab:sensitivity_analysis}. We create 14 models with varied demand from C-B (changing from 2 to 40 pedestrians) and a constant demand from B-A (8 pedestrians). Each link in the network is 2m wide with a free flow speed of 1.4 m/s. We adopt the entropy measure in the context of traffic assignment and route choice defined as

\begin{equation} \color{black}
\hspace{90pt} H = \sum_{r \in N} \sum_{s \in N} \sum_{k \in \Pi_{rs}} f_{k}^{rs} log(f_{k}^{rs}/q_{rs})  \hspace{20pt} \forall (r,s) \in W, \forall k \in \Pi_{rs}
\label{eq:entropy}
\end{equation}
where $f_{k}^{rs}$ denotes flow on route $k$ connecting OD pair $(r,s) \in W$ and $q_{rs}$ the travel demand connecting OD pair $(r,s) \in W$. We also use the ratio of the C-D-B path flow over demand C-B and the ratio of flow over capacity of link C-D as additional measures. When demand is small (less than 6 pedestrians), symmetric pVDF exhibits higher entropy and a smaller flow in path C-D-B compared with the asymmetric pVDF. When the symmetric pVDF is used, pedestrians try to avoid route C-D-B due to its large travel time and instead take the route C-A-B. When the asymmetric pVDF is used, pedestrians predominantly choose route C-D-B resulting in a smaller entropy compared with the symmetric pVDF. When demand increases, entropy for both the symmetric and asymmetric pVDFs increases consistently. The ratio of the C-B-D path flow over demand exhibits a relatively more significant difference between symmetric and asymmetric pVDFs when demand is large but still heavily bidirectional (between 8 and 20 pedestrians). See Table \ref{tab:sensitivity_analysis}.}

\begin{table}[htp]
\begin{tabular}{ccccccc} 
\hline
\multirow{2}{*}{\begin{tabular}[c]{@{}c@{}}Demand\\  C-B (ped)\end{tabular}} & \multicolumn{2}{c}{Entropy} & \multicolumn{2}{c}{Ratio of path flow C-D-B over demand} & \multicolumn{2}{c}{Flow/Capacity} \\
 & Symmetric & Asymmetric & Symmetric & Asymmetric & Symmetric & Asymmetric \\ \hline
2 & 1.27 & 0.82 & 0.67 & 0.86 & 0.05 & 0.06 \\
4 & 2.25 & 1.80 & 0.75 & 0.83 & 0.11 & 0.12 \\
6 & 3.37 & 3.37 & 0.75 & 0.75 & 0.17 & 0.17 \\
8 & 4.50 & 5.09 & 0.75 & 0.67 & 0.22 & 0.20 \\
10 & 5.62 & 6.73 & 0.75 & 0.60 & 0.28 & 0.22 \\
20 & 13.23 & 13.85 & 0.63 & 0.52 & 0.47 & 0.39 \\
40 & 27.48 & 27.70 & 0.56 & 0.52 & 0.83 & 0.77 \\ \hline
\end{tabular}
\caption{\color{black} Results of the sensitivity analysis on demand: Comparing flow patterns when deterministic symmetric and asymmetric pVDFs are used}
\label{tab:sensitivity_analysis}
\end{table}

\FloatBarrier

\subsection{A real-world large-scale network} \label{sec:largenetwork}

In this section, we apply and compare the UE-pTAP with four different pVDFs as formulated in Section \ref{sec:pVDF} on a large-scale footpath network from Sydney, Australia. The footpath network is generated automatically from the OpenStreetMap using the algorithm described in Appendix C. This network is a part of the central business district (CBD) in Sydney covering 0.55 $km^{2}$ and consists of 3,341 nodes and 19,612 links. We estimate the total demand of 213,094 trips during morning peak for 1 hour from 413 OD pairs based on information available from local transport authorities reports \citep{transport2013}. We compare the UE-pTAP solutions using four types of pVDFs. Table \ref{tab:num2_table} shows that the total system travel time (TSTT) when symmetric pVDFs are used are higher than when asymmetric pVDFs are used. Asymmetric pVDFs generate larger total number of paths than symmetric pVDFs which naturally results in higher computational cost. Stochastic pVDFs also generate larger total number of paths than deterministic pVDFs which also result in higher computational cost.  

\begin{figure}[pos=htp]
    \centering
		\includegraphics[scale=0.7]{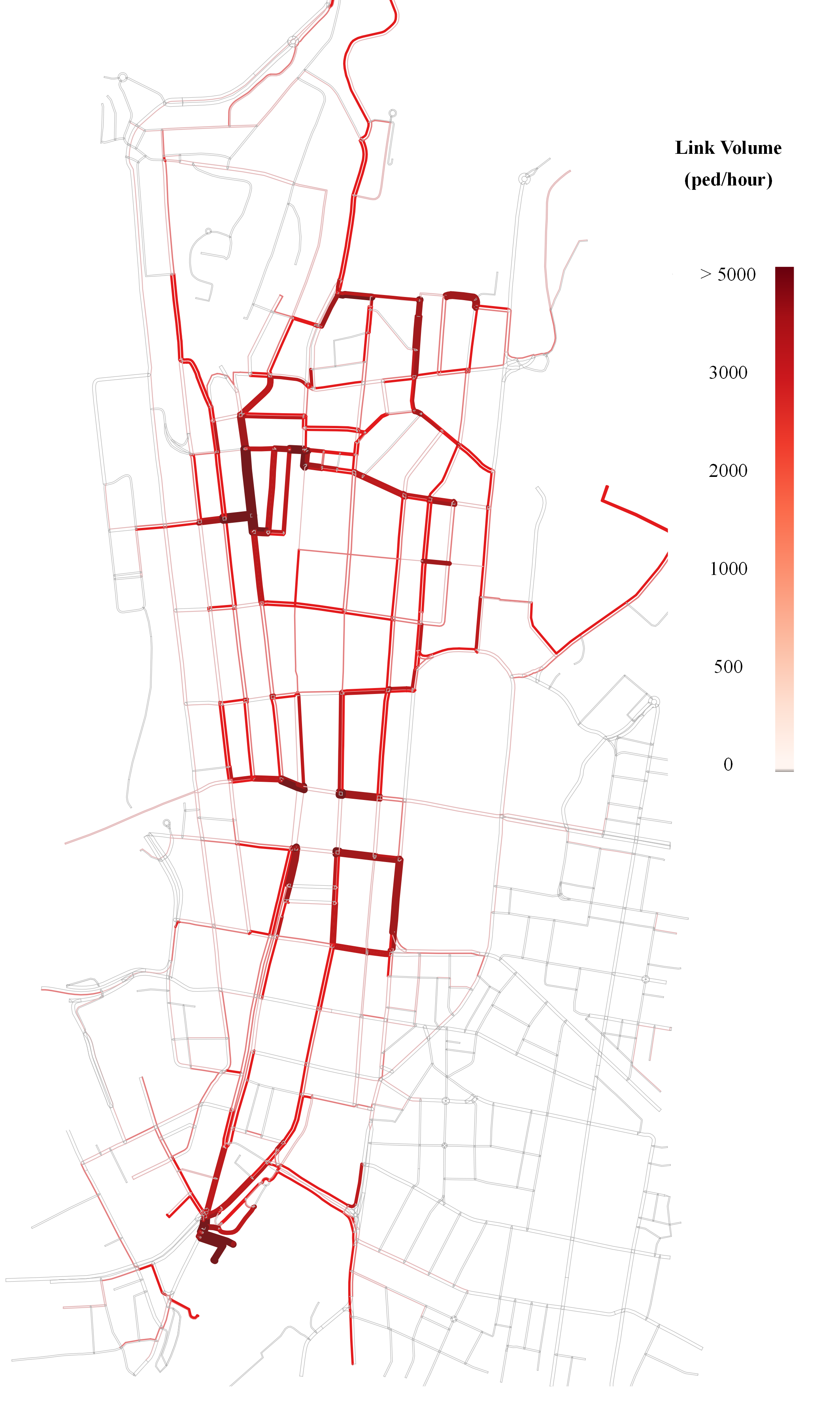}
	\caption {Illustration of estimated link volumes using the proposed UE-pTAP framework with deterministic symmetric pVDF during the morning peak (1 hour).}
    \label{fig:sydney_flowpattern}
\end{figure}

\begin{figure}[pos=htp]
    \centering
		\includegraphics[width=\textwidth]{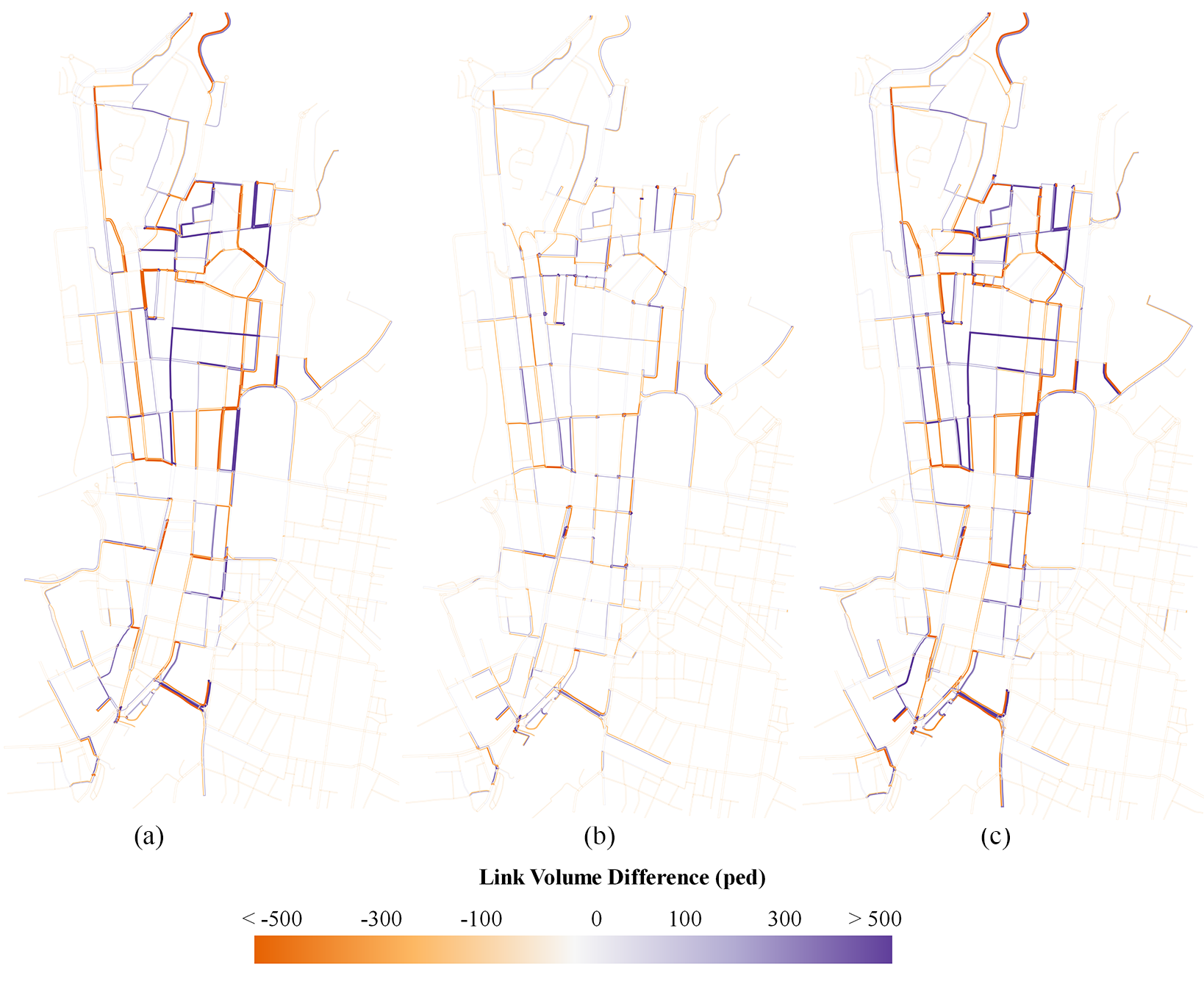}
	\caption {Illustration of link volume differences using the proposed UE-pTAP framework with different pVDFs during the morning peak (1 hour). (a) Link volume differences between deterministic asymmetric and deterministic symmetric pVDFs, (b) link volume differences between stochastic symmetric and deterministic symmetric pVDFs, and (c) Link volume differences between stochastic asymmetric and deterministic symmetric pVDFs.}
    \label{fig:sydney_linkvoldiff}
\end{figure}

\begin{figure}[pos=htp]
    \centering
    \subfloat[][]{
        \includegraphics[height = 5cm]{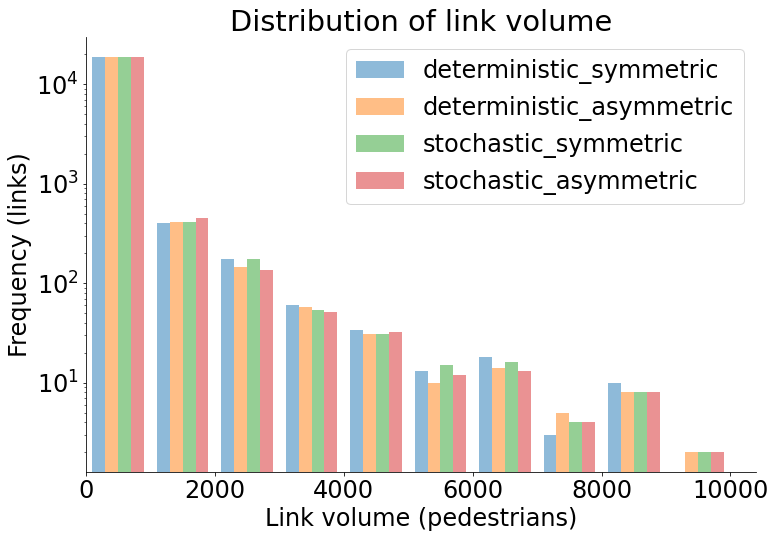}
        \label{subfig:num2_dist_linkvol}} 
    \hspace{2pt}
    \subfloat[][]{
    	\includegraphics[height = 5cm]{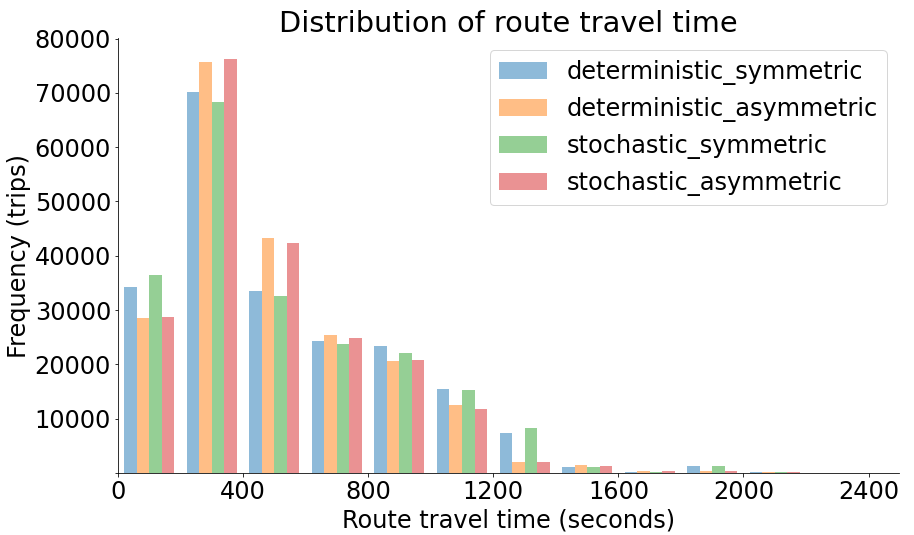}
    	\label{subfig:num2_dist_routett}} 
    
    \caption {Comparison between outputs from four pVDFs: (a) distribution of link volumes and (b) distribution of route travel times. black, orange, green, and red bars represent outputs from deterministic symmetric, deterministic asymmetric, stochastic symmetric, and stochastic asymmetric pVDFs respectively.}
    \label{fig:num2_dist}
\end{figure}

Figure \ref{fig:sydney_flowpattern} shows the estimated link volumes across the network using deterministic symmetric pVDF. Figure \ref{fig:sydney_linkvoldiff}(a)-(c) shows the link volume difference when different pVDFs are used in reference to the deterministic symmetric pVDF as previously shown in Figure \ref{fig:sydney_flowpattern}. All four types of pVDFs produce consistent link volume patterns where deterministic symmetric and stochastic symmetric pVDFs show very similar flow patterns. Deterministic asymmetric and stochastic asymmetric pVDFs show larger deviations from the deterministic symmetric pVDF. This is further confirmed in Figure \ref{fig:num2_dist} illustrating the distribution of estimated link volumes and route travel times. The most distinct differences are when route travel times are between 0 - 400 seconds as shown in Figure \ref{subfig:num2_dist_routett}. Deterministic symmetric and stochastic symmetric pVDFs share similar patterns while deterministic asymmetric and stochastic asymmetric pVDFs also follow a similar pattern. We perform Student's t-test on route travel time data and found that the observed difference between the deterministic and stochastic symmetric pVDFs is not statistically significant in the studied network. However, the difference between the outcomes of the deterministic and stochastic asymmetric pVDFs are statistically significant. This explains that the asymmetric pVDF creates greater spatial distribution of pedestrian flows during route selection. Furthermore, the observed difference between symmetric and asymmetric estimations are found to be statistically significant. We expect that the observed differences grow when congestion level in the network increases and thus, the bidirectionally effect becomes more significant. {\color{black} In this example, the difference in the performance of various pVDFs is not significant since the estimated demand level does not create high level of bidirectional pedestrian congestion. Depending on the demand level and pattern (bidirectional vs. unidirectional), the difference between symmetric and asymmetric pVDFs could be observed more clearly.}

\begin{table}[ht]
\centering 
\begin{tabular}{M c M M M M M }
\hline
pVDF type                 & TSTT (s)     & Average link volume (ped) & Total number of paths & Average path volume (ped) & Average trip travel time (s) & Number of empty links \\ \hline
Deterministic symmetric  & 111,902,157 & 131                 & 1,120                  & 60                & 525                       & 16,508              \\ 
Deterministic asymmetric & 106,396,291 & 132                 & 1,871                  & 40                & 499                       & 15,729              \\ 
Stochastic symmetric     & 112,107,692 & 134                 & 2,231                  & 27                & 526                      & 15,311              \\ 
Stochastic asymmetric    & 106,303,665 & 136                 & 2,501                  & 24                & 498                       & 14,847             \\ \hline
\end{tabular}
\caption{Comparison of network performance measures among four different types of pVDFs}
\label{tab:num2_table}
\end{table}

\subsubsection{Scenario analysis: footpath closure and walking demand}
\begin{figure}[pos=htp]
    \centering
		\includegraphics[width=1\textwidth]{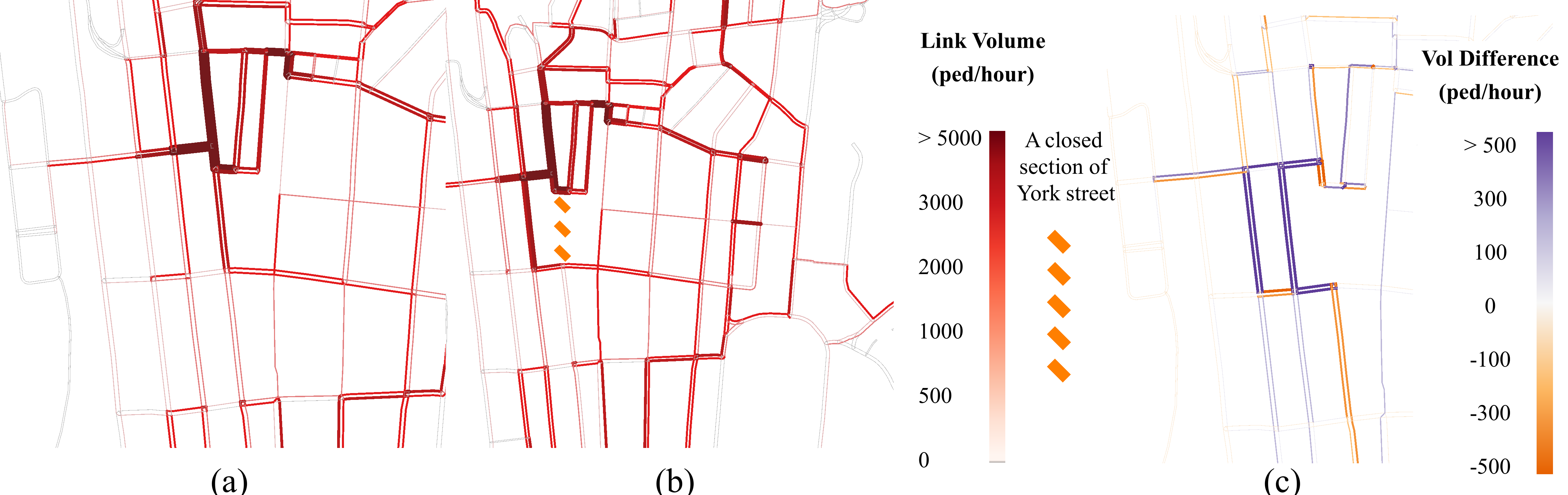}
	\caption {Application of UE-pTAP to compare the impact of footpath closure on York Street during the morning peak hour using deterministic symmetric pVDF: (a) estimated link volumes of the original network and (b) estimated link volumes of the modified network with footpath closure (c) estimated link volume difference between the modified and the original networks.}
    \label{fig:num2_closure}
\end{figure}

To further demonstrate the applicability of the model for planning purposes, we create a scenario in which a section of a busy street in Sydney CBD (York Street) is closed. We remove a section of the York Street (consisting of 4 links) from the network and perform UE-pTAP with deterministic symmetric pVDF as shown in Figure \ref{fig:num2_closure}(b). Pedestrians whose selected routes were previously crossing the closed links are now forced to find new routes. As a result, some pedestrians will also have to make detours to avoid congestion due to the new distribution of pedestrian traffic in the network. The impact of the hypothetical link closures on the the pedestrian flow distribution across the network is evident in Figure \ref{fig:num2_closure}(c). 

{\color {black} To demonstrate the impact of demand on the bidirectionality effects, we increase the original demand 10 times to 2,130,936 walking trips. The symmetric and asymmetric pVDFs capture the bidirectionality effects differently. We adopt a path flow dissimilarity measure $(\Theta _{rs})$ to quantify the relative difference between path flow patterns in the network as expressed in Equation \ref{eq:dissimilarity}. A dissimilarity measure closer to 1 suggests that path flow patterns are significantly different. While a dissimilarity measure closer to 0 suggests that the path flow patterns are similar.}

\begin{equation} \color{black}
\hspace{90pt} \Theta_{rs} = \frac{\sum _{k} ^{\Pi_{rs}} | f_{k}^{sym} - f_{k}^{asym} | }{2 q_{rs}}  \hspace{20pt} \forall (r,s) \in W
\label{eq:dissimilarity}
\end{equation}

\begin{figure}[pos=htp]
    \centering
    \includegraphics[width=0.7\textwidth]{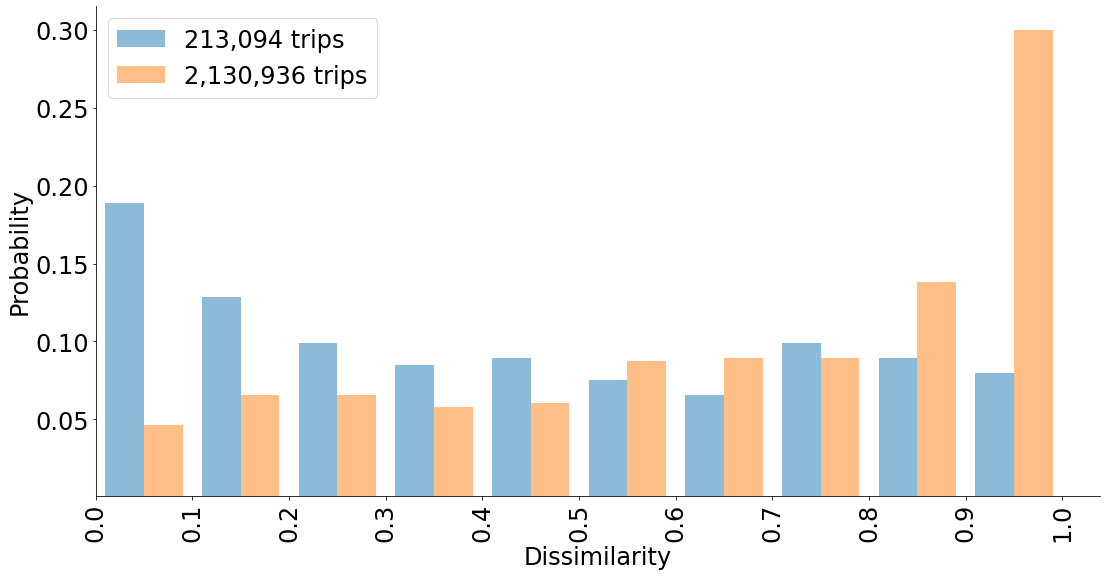}
    \caption {\color{black} Comparison between distributions of path flow dissimilarity when symmetric and asymmetric pVDFs are used under normal demand  and 10x demand.}
    \label{fig:sydney_closure_dissimilarity}
\end{figure}

{\color{black}
where $f_{k}^{sym}$ and $f_{k}^{asym}$ denote flows on route $k \in \Pi_{rs}$ from deterministic symmetric and deterministic asymmetric pVDFs respectively, $q_{rs}$ denotes demand between the OD pair r-s.

From Figure \ref{fig:sydney_closure_dissimilarity}, the distribution of the path flow dissimilarities, when symmetric vs. asymmetric pVDF is used, demonstrates the impact of demand on the bidirectionality effects and the extent the proposed pVDFs capture it differently. When demand is 10x larger, the path flow dissimilarity distribution is skewed towards left with a dominant peak at $(\Theta _{rs})$=1. While for the original demand case, the path flow dissimilarity distribution is skewed towards right with a dominant peak at $(\Theta _{rs})$=0. This means, when demand is 10x larger, many of the OD pairs have a complete path flow dissimilarity when symmetric vs. asymmetric pVDF is used.}

\section{Conclusions} \label{sec:conclusion}
This study has proposed a new user equilibrium traffic assignment framework for pedestrian networks, termed as UE-pTAP. The developed pTAP framework takes into account the impact of self-organisation in bidirectional walking streams through different types of pVDFs including symmetric versus asymmetric pVDFs and deterministic versus stochastic pVDFs. The proposed pVDFs were calibrated against controlled experimental data. Existence and uniqueness of the UE-pTAP solution is guaranteed when a symmetric pVDF is used. However, a solution for the UE-pTAP with asymmetric pVDF is not guaranteed to exist and to be unique because the asymmetric pVDF is not monotonically increasing. The UE-pTAP solution using the stochastic pVDFs provides a more realistic representation of the walking network dynamics than the deterministic pVDFs, consistent with what has been shown previously in the literature in the context of stochastic speed-flow relationships \citep{nikolic2016probabilistic}.

The proposed UE-pTAP framework has been applied on both a small-scale toy network and a large-scale real-world network to demonstrate the impact of bidirectionality of walking streams on the assignment solution and how it is captured in the proposed UE-pTAP. We showed that while all the four types of pVDFs produce different individual link volumes in the large-scale network, the overall outcomes across the network are consistent. {\color{black}The presented modeling framework in this study is expected to assist transportation authorities to better understand and measure the impact of network modification on foot traffic patterns. The model can also be used for longer term planning of walking infrastructure to accommodate the increase or decrease in the pedestrian demand due to land use change.}

A number of research directions has remained  unexplored including {\color{black}application of more advance TAP methods such as OBA \citep{bar2002origin}, LUCE \citep{gentile2014local} or TAPAS \citep{bar2010traffic},} extension of the proposed UE-pTAP to capacitated pedestrian networks and with queuing spillback, inclusion of realistic node delay functions at the signalised crossings and intersections, and a more aspirational extension to pedestrian DTA based on cell transmission model (CTM) or link transmission model (LTM). Since pedestrian dynamics evolve very quickly in reality, using hourly static pTAP modelling would not be able to capture dynamic occurrences of many microscopic behaviours. Therefore, a development of a DTA framework specifically designed and formulated for bidirectional walking networks is quite timely and needed.

\clearpage

\appendix
\section*{Appendices}
\addcontentsline{toc}{section}{Appendices}
\renewcommand{\thesubsection}{\Alph{subsection}}

\subsection{Nomenclature}\label{sec:appendix_notation}
\vspace{10pt}
\begin{center}

\begin{tabular}{ c l }
\hline

Symbols & Definition  \\ \hline
$N$ & set of all nodes in the network \\
$A$ & set of all links in the network \\
$G(N,A)$ & network graph \\

$W$ & set of origin-destination (OD) pairs \\
$R$ & set of origin nodes \\
$S$ & set of destination nodes \\
$\mathbf{x}$ & vector of all link flows \\
$\mathbf{x}^{*}$ & optimized vector of all link flows \\
$\mathbf{t}$ & vector of all link travel times \\
$c_{a}$ & flow capacity of link $a$ \\
$t_{a}$ & travel time of link $a$ \\
$\tau _{a}$ & free flow travel time of link $a$ \\
$x_{a}$ & flow on link $a$ \\
$\Pi_{rs}$ & set of all paths connecting origin nodes $r \in R$ and destination nodes $s \in S$\\
$\delta^{rs}_{a,k}$ & \makecell[l]{incident matrix indicating whether link $a$ is on path $k$  \\ between OD pair r-s or not}  \\
$f^{rs}_{a,k}$ & flow on path $k$ connecting OD pair r-s \\
$q_{rs}$ & demand between the OD pair r-s \\
$u_{i}$ & speed of data point $i$\\
$k_{i}$ & density of data point $i$ \\
$k_{c}$ & critical density \\
$c$ & flow capacity \\
$\hat{x} _{i}$ & quasi-density of data point $i$

 \\ \hline
\end{tabular}

\end{center}

\subsection{Breakdown of the Asymmetric pVDF}
\renewcommand{\figurename}{Figure \Alph{subsection}\!\!}
\setcounter{figure}{0}

\begin{figure}[pos=htp]
\centering
\includegraphics[width=0.955\linewidth]{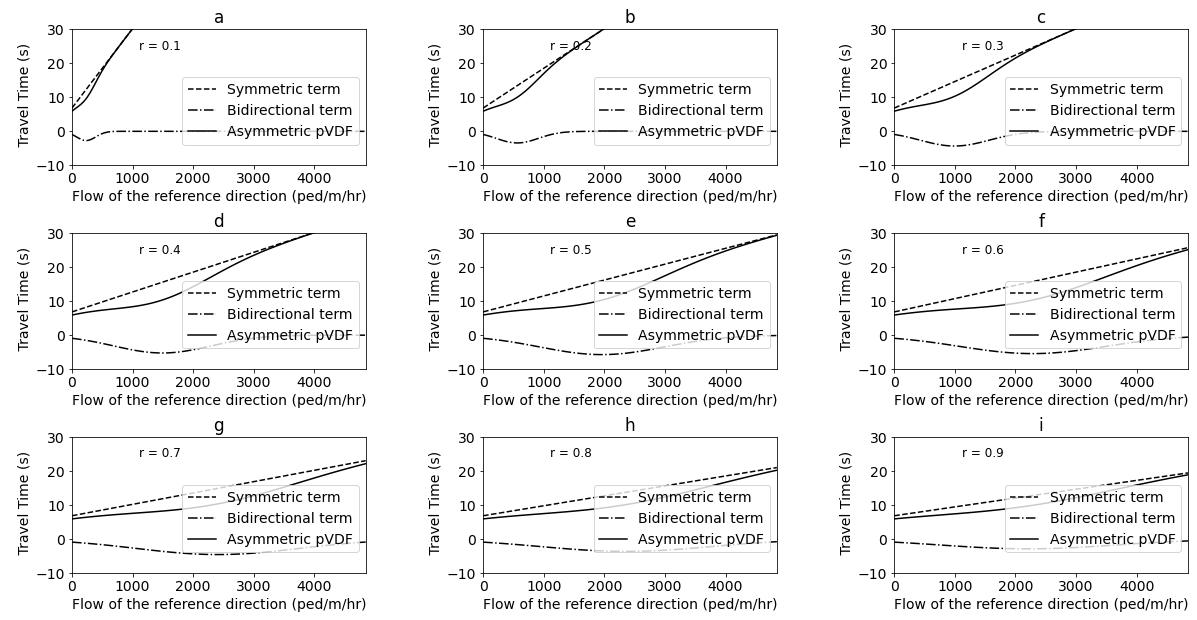}
\caption{The breakdown of the deterministic asymmetric pVDF shows the relationship between travel time and flow of the reference direction across various flow ratios. }
\label{fig:deterministic_vdf_asym_breakdown}
\end{figure}
\FloatBarrier

\subsection{Footpath network generation} \label{sec:footpathgen}
The algorithm converts a road network to pedestrian footpath network consisting of links and nodes for pTAP applications. The road network is the main input that can be obtained from OpenStreetMap (OSM). OSM is a well-known collaborative geospatial data source that includes information on road networks, points of interest, buildings, and etc. We first simplify the road network by removing unconnected lines and merging segments of the same road sections. The first step in the algorithm is to offset roads as shown in Table \eqref{tab:footpathgen_table} to create footpath links on the left and right side of the road links. The second step is to add terminal and intersection nodes. External centroids represent nodes on boundary of the network. Intersection nodes represent pedestrian crossings at the intersections. Four intersection nodes are created on every intersection. The third step is to convert footpaths into shorter footpath links. Each link is split to smaller segments based its characteristics and location of crossings. Each link contains additional information on its length, upstream node and downstream node. The fourth step is to add block centroids and four additional nodes that connect block centroids to the existing network. A midpoint node will also be created on every link that is longer than 12m. Each midpoint node will be classified into four node types including connector east, connector west, connector south or connector north based on its closet midpoint node. A block is defined by enclosing all four types of mid block nodes together to create a block centroid. The fifth step is to add connector links. Each link that contains any of the four mid block nodes will be split into new links that contain new upstream and downstream nodes. At each connector block, four connector links are created to connect block centroid nodes to all of the connectors, hence connecting block centroids to the existing network. The sixth step is to add mirror links to accommodate bidirectional effect on every footpath. Figure \ref{fig:footpathgenschematic} in the main text illustrates the output of the footpath generation algorithm.

\renewcommand{\thetable}{\Alph{subsection}\arabic{table}}
\setcounter{table}{0}
\begin{table}[ht]
\begin{center}
    \caption{Footpath network generation framework}
    \label{tab:footpathgen_table}
    \begin{tabular}{c l} 
         \hline
         
         Step 1 & Offset each road section from a road network into left and right footpath sections \\
         Step 2 & Add external centroids and intersection nodes \\
         Step 3 & Split each footpath section into links based on intersections \\
         Step 4 & Add four nodes at each block as mid block and one block centroid \\
         Step 5 & Split links and connect footpath links to connector links to block centroid \\
         Step 6 & Add mirror links for bidirectional streams \\
         \hline
     \end{tabular}
 \end{center}
\end{table}

\section*{Acknowledgement}
This research was funded by the Australian Government through the Australian Research Council (project number DP220102382).

\bibliographystyle{cas-model2-names}

\bibliography{reference}

\end{document}